\documentclass[superscriptaddress,showpacs,prb,reprint]{revtex4-1}

\usepackage{hyperref}
\usepackage{graphicx}
\usepackage{color}
\usepackage{amsbsy}
\usepackage{pstricks}
\usepackage{multirow}
\usepackage{amsmath}
\usepackage{lipsum}
\usepackage{array}
\usepackage{rotating}

\begin{document}

\title{Geometric and compositional influences on spin-orbit induced circulating currents in nanostructures}

\date{\today}

\author{J. van Bree}
\email{j.v.bree@tue.nl}
\author{A. Yu. Silov}
\author{P. M. Koenraad}
\affiliation{PSN, COBRA Research Institute, Eindhoven University of Technology, 5600 MB Eindhoven, The Netherlands}
\author{M. E. Flatt{\'e}}
\email{michael\_flatte@mailaps.org}
\affiliation{PSN, COBRA Research Institute, Eindhoven University of Technology, 5600 MB Eindhoven, The Netherlands}
\affiliation{Department of Physics and Astronomy and Optical Science and Technology Center, University of Iowa, Iowa City, Iowa 52242, USA}

\begin{abstract}
Circulating orbital currents, originating from the spin-orbit interaction, are calculated for semiconductor nanostructures in the shape of spheres, disks, spherical shells and rings for the electron ground state with spin oriented along a symmetry axis. The currents and resulting orbital and spin magnetic moments, which combine to yield the effective electron $g$ factor, are calculated using a recently introduced formalism that allows the relative contributions of different regions of the nanostructure to be identified. 
For all these spherically or cylindrically symmetric hollow or solid nanostructures, independent of material composition and whether the boundary conditions are hard or soft, the dominant orbital current originates from intermixing of valence band states in the electron ground state, circulates within the nanostructure, and peaks approximately halfway between the center and edge of the nanostructure in the plane perpendicular to the spin orientation. For a specific material composition and confinement character, the confinement energy and orbital moment are determined by a single size-dependent parameter for spherically symmetrical nanostructures, whereas they can be independently tuned for cylindrically symmetric nanostructures.
\end{abstract}

\maketitle

\section{Introduction}

Spin-correlated orbital currents provide the source\cite{Bree2014} of the dramatic modification of the effective magnetic moment ${\boldsymbol \mu}$ of the electron in a semiconductor\cite{Roth1959,Yafet1963}. Confinement has been shown to quench this magnetic moment, even for nanostructures with spherical symmetry\cite{Pryor2006b,Pryor2007,Bree2012,Bree2014}, 
to a much greater degree than expected from confinement-induced shifts in semiconductor band gap, spin-orbit splitting, and masses. Confinement-induced effects on the magnetic moment ${\boldsymbol \mu}$ also directly modify the temporal evolution of a spin in a magnetic field\cite{Vrijen2000,Jiang2001,Xiao2004,Nakaoka2007,Pingenot2008,De2009,Andlauer2009,Pingenot2011,Roloff2010}, by slowing or speeding precession, or through forms of electrically-driven resonance such as $g$ tensor modulation resonance\cite{Kato2003}.
These modifications have been suggested as means to manipulate the spins for quantum computation\cite{Loss1998,Kane1998}. Recently the spatial structure of these orbital currents were calculated directly in spherical and cylindrical III-V semiconductor nanostructures\cite{Bree2014}
 and the peak currents were identified to be midway from the center of the nanostructure to the edge of the nanostructure in the plane perpendicular to the magnetic moment's orientation. This suggests that removing the material in the center of the nanostructure, forming a shell or ring, might have minimal effect on the electron's magnetic moment. It also suggests where electrical gates might be positioned to have the greatest effect on the electron's magnetic moment.

Here we calculate the spin-correlated orbital currents for spheres, cylinders, spherical shells and rings, identifying the response of the spin-correlated orbital currents to changes in topology, to changes in disk and ring aspect ratio, and to the softness of the confining potential. The overall conclusions of Ref.~\onlinecite{Bree2014} regarding the source of the orbital current remain valid in these structures. That is, the dominant orbital contribution to the spin's magnetic moment originates from a ground-state, dissipationless current loop circulating within the dot. The calculations use semiconductor envelope-function theory for direct-gap semiconductor quantum dots\cite{Vahala1990,Sercel1990}. The contributions from spin-orbit-correlated circulating currents are fully identified and broken down into constituent contributions. Contributions largely neglected in Ref.~\onlinecite{Bree2014} because they are not the largest contributors to the magnetic moment include contributions from orbital currents within a unit cell and contributions associated with a single envelope function; both are discussed in detail here. The boundary conditions for these nanostructures are considered to be hard-wall, which are appropriate for many colloidal quantum dots and nanowires, or harmonic and soft, characteristic of electrostatic confinement. Although this approach can, in principle, be generalized to other electronic states, including excited electronic states and hole states, this generalization requires dealing with significant additional complexities associated with non-zero angular momentum in the conduction-band envelope functions. Thus here we focus on orbital contributions to the magnetic moment along a symmetry axis of a sphere, shell, disk or ring; in-plane electron magnetic moments will be the subject of future work.

The paper's structure is as follows. In Sec.~\ref{sec:frame} the theoretical formalism  introduced in Ref.~\onlinecite{Bree2014} to calculate the orbital contributions to the spin's magnetic moment is summarized, as it is relied on for later sections. The formalism is then applied to spheres in Sec.~\ref{sec:spheres}, spherical shells in Sec.~\ref{sec:shells}, disks with hard-wall boundaries in Sec.~\ref{sec:harddisks}, disk with soft boundaries in Sec.~\ref{sec:softdisks}, and rings in Sec.~\ref{sec:rings}. We finally draw general conclusions on all these different geometries in Sec.~\ref{sec:con}.

\section{Theoretical framework}\label{sec:frame}

Throughout this article, we focus on the spin-oriented electron ground state $\Psi({\bf r})$ of a nanostructure. The magnetic moment ${\boldsymbol \mu}_{\text{tot}}$  contains contributions from both the spin and the orbital motion of the state:
\begin{eqnarray}
{\boldsymbol \mu}_{\text{tot}} = {\boldsymbol \mu}_{\text{spin}} + {\boldsymbol \mu}_{\text{orb}}
\end{eqnarray}
This moment can couple to an external applied magnetic field ${\bf B}$ via the Zeeman interaction:
\begin{eqnarray}
{\cal H}_{\text{Zeeman}} = -{\boldsymbol \mu}_{\text{tot}} \cdot {\bf B}
\end{eqnarray}
In absence of a magnetic field, the ground state will be  degenerate due to time-reversal invariance\cite{Kramers1930}; two degenerate states are the time reversal of each other, and have an oppositely oriented magnetic moment. It therefore suffices to examine only one state of the Kramers doublet. In this article we fix the orientation of the magnetic moment along the symmetry axis of the nanostructure, which can be experimentally realized by either electrical spin injection or optical orientation.

The magnetic moment is related to the $g$-factor, which is often used in an experimental context and can be defined as\cite{Pryor2007}:
\begin{eqnarray}
g = \frac{E_{\uparrow}-E_{\downarrow}}{\mu_B B},
\end{eqnarray}
where $E_{\uparrow,\downarrow}$ are the energies associated with spin up/down, and $\mu_B={e\hbar}/{2 m_0}$ is the Bohr magneton. Using the Zeeman interaction and time-reversal symmetry, we can relate the $g$-factor to the magnetic moment in the limit of zero magnetic field:
\begin{eqnarray}
|g| &=& \lim_{B\rightarrow 0} \frac{(-{\boldsymbol \mu}_{\text{tot}}\cdot{\bf B})-({\boldsymbol \mu}_{\text{tot}}\cdot{\bf B})}{\mu_B B} \\
&=& 2 \frac{\mu_{\text{tot}}}{\mu_B} = 2\left(\frac{\mu_{\text{spin}}}{\mu_B} + \frac{\mu_{\text{orb}}}{\mu_B}\right)
\end{eqnarray}
where we assumed the magnetic field and the magnetic moment to be collinear. We would like to stress that the factor $2$ has no relation to the free electron $g$-factor, and stems solely from the Kramers degeneracy. The term $\mu_{\text{orb}}$ refers to the orbital contribution to the {\it spin}'s magnetic moment, not the spin-independent orbital moment of the electron. Subsequently in the text, however, we will  refer to this  simply as the orbital magnetic moment. Even though we will focus on the spin and orbital contributions to the spin's magnetic moment in the rest of this article, the above relation enables us to connect them to an experimentally measurable $g$-factor. In the next two sections we present the theoretical framework to calculate the orbital and spin moments.

\subsection{Orbital moment} \label{sec:orbmu}

\begin{figure}
\begin{center}
\includegraphics[width=0.8\columnwidth]{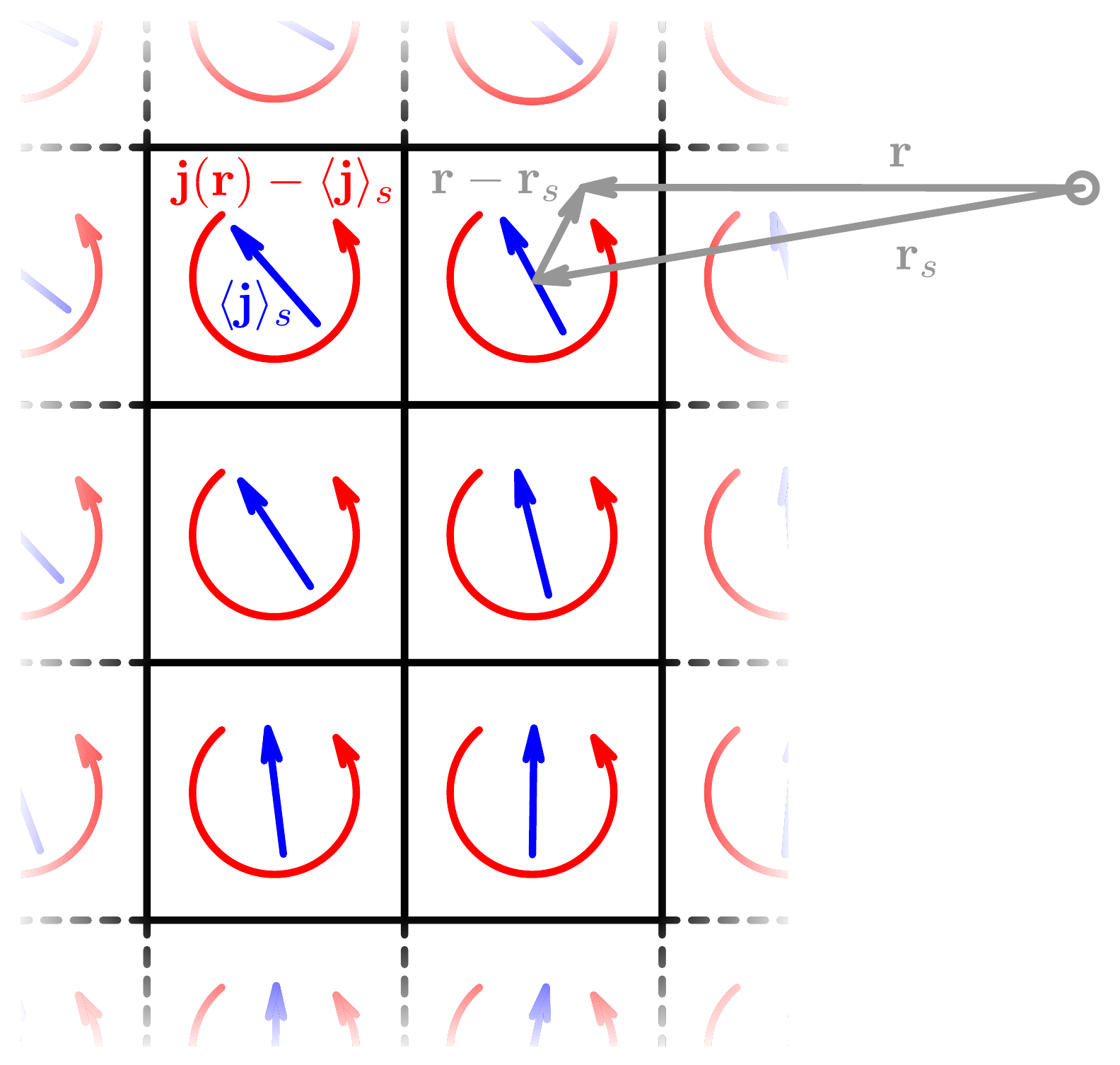}
\caption{The orbital current within a unit cell can be split into an itinerant contribution $\langle{\bf j}\rangle_s$, and a localized contribution ${\bf j}({\bf r})-\langle{\bf j}\rangle_s$. Vector ${\bf r}_s$ points to the center of unit cell $s$.}
\label{fig:unit_cell}
\end{center}
\end{figure}

The orbital magnetic moment ${\boldsymbol \mu}_{\text{orb}}$ is related\cite{Jackson1998b} to the orbital current density ${\bf j}\left({\bf r}\right)$ by
\begin{eqnarray}
{\boldsymbol \mu}_{\text{orb}} &=& \frac{1}{2}\int_V{\bf r}\times{\bf j}\left({\bf r}\right)~d^3r = \frac{1}{2} \sum_s \int_{V_s} {\bf r} \times {\bf j}({\bf r})~d^3r,
\end{eqnarray}
where we have considered the moment as a summation of moments arising from each of $s$ unit cells having volume $V_s$. We define the average current density $\langle {\bf j} \rangle_s$ in a unit cell as:
\begin{eqnarray}
\langle {\bf j} \rangle_s = \frac{1}{V_s} \int_{V_s} {\bf j}({\bf r})~d^3r.
\end{eqnarray}
Using $\langle{\bf j}\rangle_s$ we  split the orbital current into an itinerant current (IC) that flows into and out of a unit cell, and a localized current (LC) whose average over the unit cell vanishes, given by ${\bf j}({\bf r}) - \langle{\bf j}\rangle_s$ (see also Fig.~\ref{fig:unit_cell}). The magnetic moment can then be expressed as\cite{Thonhauser2005}:
\begin{eqnarray}
{\boldsymbol \mu}_{\text{orb}} &=& \frac{1}{2}\sum_{s} \Big\{ \underbrace{V_s {\bf r}_s \times \langle {\bf j} \rangle_s}_{\text{Itinerant current (IC)}} \nonumber \\
 && \quad \quad \quad + \underbrace{\int_{V_s} ({\bf r} - {\bf r}_s) \times \left\{{\bf j}({\bf r}) - \langle {\bf j} \rangle_s\right\}~d^3r}_{\text{Localized (circulating) current (LC)}} \Big\}
\end{eqnarray}
where ${\bf r}_s$ is the vector pointing to unit cell $s$. The first term is the orbital moment due to itinerant currents, while the second term is the sum of orbital moments due to a (circulating) current localized within each unit cell. For an isolated atom, the first term is zero. The spatial extent of states in semiconductors can be substantial, leading to a much larger lever arm for the moments arising from itinerant currents than for the moments arising from localized currents (i.e. ${\bf r}_s \gg {\bf r} - {\bf r}_s$). These orbital currents follow from\cite{Messiah1961}:
\begin{eqnarray}
{\bf j}\left({\bf r}\right) = \frac{e\hbar}{m_0} \text{Im}\left\{\Psi^*\left({\bf r}\right)\nabla\Psi\left({\bf r}\right)\right\}
\end{eqnarray}

\noindent
One can conceptually think of the state $\Psi({\bf r})$ as being composed out of traveling and standing waves. The latter are formed by reflection from the boundaries of the nanostructure, whereas the former are contained inside the nanostructure. The current is therefore directed parallel to the surface of the nanostructure and is divergence-free, which is in compliance with the assertion of $\Psi({\bf r})$ being a stationary state. To understand the origin of the traveling waves, we need to describe the state $\Psi\left({\bf r}\right)$ in more detail. For nanostructures, the envelope function approximation (EFA) is an accurate way to do so\cite{Luttinger1955,Luttinger1956}:
\begin{eqnarray}
\Psi({\bf r}) = \sum_i F_i({\bf r}) u_i({\bf r}),
\end{eqnarray}
where the wave function is written as the product of a Bloch state $u_i({\bf r})$ of band $i$ with its corresponding spatially slowly varying envelope function $F_i({\bf r})$, assumed to be constant in a unit cell. This results in currents which are related to the Bloch velocity (BV) and envelope velocity (EV):
\begin{eqnarray}
{\bf j}({\bf r}) = \frac{e\hbar}{m_0} \sum_{i,j} &\text{Im}& \Big\{ \underbrace{u_i^*({\bf r}) u_j({\bf r}) \left[F_i^*({\bf r}) \nabla F_j({\bf r})\right]}_{\text{Envelope velocity related (EV)}} \nonumber \\
&&+ \underbrace{F_i^*({\bf r}) F_j({\bf r}) \left[u_i^*({\bf r}) \nabla u_j({\bf r})\right]}_{\text{Bloch velocity related (BV)}} \Big\}
\end{eqnarray}
The BV related current dominates over the EV related current by $\sim d/a_0$, where $d$ is the typical size of the envelope wave function and $a_0$ the size of the unit cell\cite{Bree2014}. This coincides with the condition for the validity of the envelope function approximation. For realistically sized nanostructures, the BV related current is $\geq5$ times the EV related current. For illustrative purposes consider the states labeled by $i$ as originating from the conduction band, whereas states labeled by $j$ are related to the valence band. It is then apparent that almost all of the more important BV related orbital current arises due to intermixing of valence states into the electron ground state of a nanostructure; the $F_j({\bf r})$ must be non-zero. The minimal model to accurately calculate the orbital current must therefore contain at least the eight bands describing the conduction and valence band. We will now separately work out the BV and EV related currents.

The unit cell averaged current density for the BV related current $\langle{\bf j}\rangle^{\text{BV}}$ becomes:
\begin{eqnarray}
\langle {\bf j} \rangle^{\text{BV}}({\bf r}_s) = \frac{1}{V_s} \frac{e\hbar}{m_0} \sum_{i,j} \text{Im}\left\{ F^*_i({\bf r}_s)F_j({\bf r}_s) \langle u_i | \nabla | u_j \rangle \right\} \hspace{6mm} \label{eq}
\end{eqnarray}
where $\langle u_i | \nabla | u_j \rangle$ are momentum matrix elements. These are only non-zero when $i$ labels a conduction band state and $j$ a valence band state. For the electron ground state in a nanostructure, $F_i({\bf r}_s)$ will be an $s$-like state and $F_j({\bf r}_s)$ a $p$-like state. The product of these envelope wave functions will therefore peak roughly midway between the center and edge of a nanostructure. Since we are examining a stationary state, the divergence of the current is zero. The current must therefore circulate within the nanostructure along a closed surface. This resembles a current loop extended throughout the nanostructure and arising completely from intermixing of valence band states in the ground state of the nanostructure. This BV related itinerant current leads to a magnetic moment 
\begin{eqnarray}
{{\boldsymbol \mu}_{\text{IC-BV}}({\bf r}_s)} = \mu_B\sum_{i,j} \text{Im} \{ F^*_i({\bf r}_s)F_j({\bf r}_s) ({\bf r}_s \times \langle u_i | \nabla | u_j \rangle) \}. \hspace{7mm} \label{eq:mu_ICBV}
\end{eqnarray}
The BV related localized current leads to a magnetic moment 
\begin{eqnarray}
{{\boldsymbol \mu}_{\text{LC-BV}}({\bf r}_s)} = {\mu_B}\sum_{i,j} \text{Im} \{ F^*_i({\bf r}_s)F_j({\bf r}_s) \langle u_i | {\bf L}_{\text{B}} | u_j \rangle \}, \hspace{3mm} \label{eq:mu_LCBV}
\end{eqnarray}
where ${\bf L}_{\text{B}}=({\bf r}-{\bf r}_s)\times\nabla$ is the angular momentum operator acting on the Bloch functions. The Bloch angular momentum does not exceed $1$, and therefore ${\boldsymbol \mu}_{\text{IC-BV}}\gg{\boldsymbol \mu}_{\text{LC-BV}}$. Whereas the spatial distribution of ${\boldsymbol \mu}_{\text{IC-BV}}$ follows that from the above-discussed $\langle{\bf j}\rangle_{\text{BV}}$, the spatial distribution of ${\boldsymbol \mu}_{\text{LC-BV}}$ is given by the the product of two $p$-like envelope states, since the Bloch angular momentum is only non-zero for valence states. The spatial distributions of both magnetic moments have therefore an odd spatial symmetry.

\begin{table*}
\caption{Material parameters of the different zinc-blende materials used throughout the article.}
\begin{ruledtabular}
\begin{tabular}{r|cccccccccc}
Material & $E_g$ [eV] & $\Delta$ [eV] & $P_0$ [eV\AA] & $\gamma_1^L$ & $\gamma_2^L$ & $\gamma_3^L$ & $|\chi|$ & $\mu_{\text{Roth}}$ [$\mu_B$] & $\mu_{\text{orb,exp}}$ [$\mu_B$]\cite{Madelung2004} & Deviation of $\mu$ [$\mu_B$]\\
\hline
HgTe\cite{Lawaetz1971}    & -0.303 & 1.00  & 8.29  & -18.68 & -10.19 & -9.56 & 31.1 & $|28.4|$& $|21.5|$& 6.9 (+32\%) \\
InSb\cite{Vurgaftman2001} & 0.235  & 0.81  & 9.426 & 34.8   & 15.5   & 16.5  & 32.2 & -25.6 & -26.7 & 1.1 (-4\%) \\
InAs\cite{Vurgaftman2001} & 0.417  & 0.39  & 9.055 & 20     & 8.5    & 9.2   & 25.5 & -8.19 & -8.65 & 0.46 (-5\%) \\
Ga$_{0.47}$In$_{0.53}$As\cite{Vurgaftman2001} & 0.816 & 0.329 & 9.47 & 11.01 & 4.18 & 4.84 & 13.9 & -2.76 & -3.25 & 0.49 (-15\%) \\
GaAs\cite{Vurgaftman2001} & 1.519  & 0.341 & 9.764 & 6.98   & 2.06   & 2.93  & 5.9  & -1.00 & -1.22 & 0.22 (-18\%) \\
CdTe\cite{Lawaetz1971}    & 1.60   & 0.91  & 8.88  & 5.29   & 1.89   & 2.46  & 7.8  & -1.56 & -1.83 & 0.27 (-15\%) \\
CdSe\cite{Kim1994}        & 1.84   & 0.42  & 7.40  & 3.38   & 1.12   & 1.47  & 7.6  & -0.48 &       &  \\
ZnTe\cite{Lawaetz1971}    & 2.39   & 0.92  & 8.53  & 3.74   & 1.07   & 1.64  & 5.0  & -0.74 & -1.21 & 0.47 (-38\%) \\
ZnSe\cite{Lawaetz1971}    & 2.82   & 0.43  & 9.61  & 3.77   & 1.24   & 1.67  & 7.0  & -0.38 & -0.47 & 0.09 (-19\%) \\
ZnS\cite{Lawaetz1971}     & 3.80   & 0.07  & 8.82  & 2.54   & 0.75   & 1.06  & 5.6  & -0.03 & -0.06 & 0.03 (-50\%) \\
\end{tabular}
\label{table:materials}
\end{ruledtabular}
\end{table*}

The unit cell averaged current density for the EV related current 
\begin{eqnarray}
\langle {\bf j} \rangle^{\text{EV}}({\bf r}_s) = \frac{1}{V_s} \frac{e \hbar}{m_0} \sum_{i} \text{Im} \{ F^*_i({\bf r}_s) \nabla F_i({\bf r}_s)\}, \label{eq:mu_ICEV}
\end{eqnarray}
where we have used the orthonormality of the Bloch functions. The envelope wave function of the conduction band does not contribute to this current for the electron ground state, since it does not consist of a traveling wave. This current is therefore solely determined by the envelope wave functions associated with the valence band; the spatial distribution of $\langle {\bf j} \rangle_{\text{EV}}$ is the product of a $p$-like $F_i({\bf r}_s)$ and a $p$-like $\nabla F_i({\bf r}_s)$, and has therefore the same odd spatial symmetry as $\langle{\bf j}\rangle_{\text{BV}}$. The magnetic moment ${\boldsymbol \mu}_{\text{IC-EV}}$ originating from the EV related itinerant current becomes:
\begin{eqnarray}
{{\boldsymbol \mu}_{\text{IC-EV}}({\bf r}_s)} = {\mu_B}\sum_{i} \text{Im} \{ F^*_i({\bf r}_s) {\bf L}_{\text{E}} F_i({\bf r}_s)\} ,\label{eq:mu_LCEV}
\end{eqnarray}
where ${\bf L}_{\text{E}} = {\bf r}_s \times \nabla$ is the angular momentum operator acting on the envelope wave functions. The EV-related localized current leads to a magnetic moment 
\begin{eqnarray}
{{\boldsymbol \mu}_{\text{LC-EV}}({\bf r}_s)} = {\mu_B}\sum_{i,j} \text{Im} \{ F^*_i({\bf r}_s) \langle u_i | {\bf r} - {\bf r}_s | u_j \rangle \times \nabla F_j({\bf r}_s) \} \nonumber \\
\label{eq:LCEV}
\end{eqnarray}
where $\langle u_i | {\bf r} - {\bf r}_s | u_j \rangle$ are dipole matrix elements. These are only non-zero when $i$ labels a conduction band state and $j$ a valence band state, because of the parity quantum numbers. This means that the spatial distribution will have an even spatial symmetry: both $F_i({\bf r}_s)$ and $\nabla F_j({\bf r}_s)$ are $s$-like. This is different from the other contributions to the orbital moment, which all have an odd spatial symmetry. We can relate the dipole matrix elements to the momentum matrix elements through the commutation relation\cite{Gu2013}:
\begin{eqnarray}
[{\cal H},{\bf r}] = \frac{\hbar}{i m_0}{\bf p}
\end{eqnarray}
by which
\begin{eqnarray}
\langle \phi_i | {\bf p} | \phi_j \rangle &=& \langle \phi_i | \frac{i m_0}{\hbar} [{\cal H},{\bf r}] | \phi_j \rangle\\
&=& \frac{i m_0}{\hbar} (E_i - E_j) \langle \phi_i | {\bf r} | \phi_j \rangle
\end{eqnarray}
and therefore
\begin{eqnarray}
\langle u_i | {\bf r} - {\bf r}_s | u_j \rangle = -\frac{\hbar^2}{m_0 (E_i - E_j)} \langle u_i | \nabla | u_j \rangle \label{eq:energy}
\end{eqnarray}

The total orbital moment ${\boldsymbol \mu}_{\text{orb}}$ is the sum of ${\boldsymbol \mu}_{\text{IC-BV}}$, ${\boldsymbol \mu}_{\text{LC-BV}}$, ${\boldsymbol \mu}_{\text{IC-EV}}$, and ${\boldsymbol \mu}_{\text{LC-EV}}$. We replace the summation over $s$ with an integral over the whole volume of the state, since the state $\Psi({\bf r})$ is extended over many unit cells.

\subsection{Spin moment}

Using the non-relativistic limit of the Dirac equation, we find the spin moment to be given by\cite{Messiah1961}:
\begin{eqnarray}
{\boldsymbol \mu}_{\text{spin}} = \frac{e\hbar}{2m_0} \sum_s \int_{V_s} \Psi^*({\bf r}){\boldsymbol \sigma}\Psi({\bf r})~d^3r
\end{eqnarray}
where ${\boldsymbol \sigma}=\left(\sigma_x,\sigma_y,\sigma_z\right)$ is the Pauli vector, with $\sigma_{x,y,z}$ the Pauli matrices. We have again split the integration over the whole state into a summation of integrations over the unit cell. We can then proceed and use the EFA for the wave function $\Psi({\bf r})$, by which the spin moment becomes
\begin{eqnarray}
{{\boldsymbol \mu}_{\text{spin}}({\bf r}_s)} = {\mu_B}\sum_{i,j} F_i^*({\bf r}_s)F_j({\bf r}_s) \langle u_i | {\boldsymbol \sigma} | u_j \rangle. \label{eq:muspin}
\end{eqnarray}
The spatial structure of the spin moment is therefore given by the product $F_i^*({\bf r}_s)F_j({\bf r}_s)$. If we assume that the electron ground state of the nanostructure is dominated by the conduction band state, the spatial distribution of the spin moment is approximately $|F_i({\bf r}_s)|^2$, where $F_i({\bf r}_s)$ is an $s$-like envelope wave function. This even spatial symmetry is markedly different from the odd spatial symmetry of the dominant orbital moment density ${\boldsymbol \mu}_{\text{IC-BV}}({\bf r}_s)$.

\subsection{Boundary conditions and ${\bf k}\cdot{\bf p}$-model}

As mentioned in Sec.~\ref{sec:orbmu}, any accurate calculation of the orbital current should include a minimum of eight bands. To keep the problems analytically tractable, we choose a standard eight-band ${\bf k}\cdot{\bf p}$-model\cite{Sercel1990} and hard-wall boundaries for most of the nanostructures. Within ${\bf k}\cdot{\bf p}$-theory, boundary conditions have been the subject of debate\cite{Bastard1988,Burt1992,Rodina2002}. Since we can assume that the electron ground state is dominated by conduction band states, we pragmatically opt for the approximate boundary condition that only the conduction band envelope wave function needs to vanish at the boundary. This approximation is exact for the bulk and has as much validity as hard-wall boundaries and the envelope function approximation itself.


To illustrate our analytical results, we  show numerical calculations for nanostructures of different materials with a zinc-blende crystal structure. The corresponding material parameters are tabulated in Table~\ref{table:materials}, where $E_g$ is the band gap energy, $\Delta$ the spin orbit splitting, $P_0$ the momentum matrix element, and $\gamma_{1,2,3}^L$ are the Luttinger parameters. In an eight-band ${\bf k}\cdot{\bf p}$-calculation, the Luttinger parameters need to be modified for the explicit inclusion of the $\Gamma_6^c$-band\cite{Winkler2003}:
\begin{eqnarray}
\gamma_1 = \gamma_1^L - \frac{1}{3}\frac{2 m_0}{\hbar^2}\frac{P_0^2}{E_g}\\
\gamma_2 = \gamma_2^L - \frac{1}{6}\frac{2 m_0}{\hbar^2}\frac{P_0^2}{E_g}\\
\gamma_3 = \gamma_3^L - \frac{1}{6}\frac{2 m_0}{\hbar^2}\frac{P_0^2}{E_g}
\end{eqnarray}
For most materials there is a fairly large spread in the reported values of the $\gamma_{1,2,3}^L$-parameters and $P_0$, which reflects the degree of accuracy of the ${\bf k}\cdot{\bf p}$-model. Still, the bulk orbital moment $\mu_{\text{Roth}}$ is fairly well reproduced using an eight-band model (see Table~\ref{table:materials}): the agreement is for most materials within $0.5\mu_B$ or $15-20\%$. It can clearly be observed that the model becomes less accurate as $E_g$ increases, since the remote bands become of equal importance to the eight bands that are explicitly included. An improvement of the eight band model would involve inclusion of the $\Gamma_{7,8}^c$-bands~\cite{Hermann1977}. These bands would generate additional $p$-like envelope wave functions in the electron ground state, and therefore generate similar spin-orbit correlated currents as the $\Gamma_{7,8}^v$-bands. We therefore do not expect any additional features by including additional bands, except for improving the quantitative agreement.

\section{Spherical symmetry} \label{sec:spherical}

We will first examine nanostructures having spherical symmetry. The envelope functions of such nanostructures will exhibit spherical symmetry, if both the confinement potential and the crystal have spherical symmetry. Fortunately, the anisotropy of the valence band is rather small for most semiconductors. This can be formally analyzed by decomposing the Hamiltonian into spherically and cubically symmetric terms\cite{Baldereschi1973}. The ratio of the spherical over cubic terms can be expressed as:
\begin{eqnarray}
\chi=\frac{2}{5} \left(\frac{2\gamma_2+3\gamma_3}{\gamma_3-\gamma_2}\right)
\end{eqnarray}
From Table~\ref{table:materials} we see that the spherical terms are at least 5 times larger than the cubic terms, so we can safely assume that the crystal has spherical symmetry. In the spherical approximation, the Hamiltonian will be block diagonal in a basis of eigenstates of ${\bf F}$ and $F_z$, where the total angular momentum ${\bf F} = {\bf L}_{\text{E}} + {\bf J} = {\bf L}_{\text{E}} + {\bf L}_{\text{B}} + {\bf s}$ (${\bf J}$ the total Bloch momentum, ${\bf s}$ the spin moment)\cite{Vahala1990}:
\begin{eqnarray}
{\cal H} = \sum_{F,F_z} {\cal H}_{F,F_z}.
\end{eqnarray}
These basis states can be found by using the rules for adding angular momenta,
\begin{eqnarray}
&&|F,F_z;J,L_{\text{E}};k\rangle = \\
&&\sum_{J_z=-J}^J \sum_{L_{\text{E},z}=-L_{\text{E}}}^{L_{\text{E}}} \langle J,J_z;L_{\text{E}},L_{\text{E},z} | F,F_z \rangle |J,J_z\rangle |k,L_{\text{E}},L_{\text{E},z}\rangle,\nonumber
\end{eqnarray}
where $\langle J,J_z;L_{\text{E}},L_{\text{E},z} | F,F_z \rangle$ are Clebsch-Gordan coefficients, $|J,J_z\rangle$ the Bloch functions, and $|k,L_{\text{E}},L_{\text{E},z}\rangle$ the envelope wave functions. This notation is slightly different from Sec.~\ref{sec:frame}, where Bloch functions are denoted as $u_i({\bf r})$, and envelope wave functions as $F_i({\bf r})$. The envelope wave function has the coordinate representation:
\begin{eqnarray}
&& \langle r,\theta,\phi |k,L_{\text{E}},L_{\text{E},z}\rangle = \nonumber \\
&& \quad \quad \sqrt{\frac{2}{\pi}} i^{L_{\text{E}}} \{j_{L_{\text{E}}}\left(kr\right)+ \xi y_{L_{\text{E}}}\left(kr\right) \} Y_{L_{\text{E}}}^{L_{\text{E},z}}(\theta,\phi)
\end{eqnarray}
where $j_l(r)$ is the $l$th-order spherical Bessel function of the first kind, $y_l(r)$ is the $l$th-order spherical Neumann function of the first kind, $Y_l^m\left(\theta,\phi\right)$ a spherical harmonic, and $\xi$ a dimensionless parameter determined by the boundary conditions.
For the electron ground state it suffices to examine the $|F,F_z\rangle=|\frac{1}{2},+\frac{1}{2}\rangle$ subspace, since this is the lowest possible $F$ and $|\tfrac{1}{2},-\tfrac{1}{2}\rangle$ is the time-reversed state of $|\tfrac{1}{2},+\tfrac{1}{2}\rangle$. Within an eight-band ${\bf k}\cdot{\bf p}$-model, the $|\tfrac{1}{2},+\tfrac{1}{2}\rangle$ subspace is spanned by three basis states: $|\tfrac{1}{2},+\tfrac{1}{2};\tfrac{1}{2},0;k\rangle$, $|\tfrac{1}{2},+\tfrac{1}{2};\tfrac{3}{2},1;k\rangle$, and $|\tfrac{1}{2},+\tfrac{1}{2};\tfrac{1}{2},1;k\rangle$. Following the transformation of Ref.~\onlinecite{Sercel1990}, we can represent the Hamiltonian in this basis:
\begin{eqnarray}
&&{\cal H}_{\frac{1}{2},+\frac{1}{2}} = \\
&& \left(
\begin{array}{ccc}
\frac{\hbar^2}{2 m_0}k^2 & -i\sqrt{\frac{2}{3}}P_0 k & -i\sqrt{\frac{1}{3}}P_0 k \\
i\sqrt{\frac{2}{3}}P_0k & -E_g-\frac{\hbar^2}{m_0}\frac{\gamma_1+2\gamma_{23}}{2}k^2 & -\sqrt{2}\frac{\hbar^2}{m_0}\gamma_{23}k^2 \\
i\sqrt{\frac{1}{3}}P_0k & -\sqrt{2}\frac{\hbar^2}{m_0}\gamma_{23}k^2 & -E_g-\Delta-\frac{\hbar^2}{m_0}\frac{\gamma_1}{2}k^2
\end{array}
\right)\nonumber
\end{eqnarray}
where $k$ is the radial wave number, and $\gamma_{23}=\tfrac{2}{5}\gamma_2 + \tfrac{3}{5}\gamma_3$ the modified spherical Luttinger parameters. The electron ground state can be expressed as a linear combination of the three basis states:
\begin{eqnarray}
|\Psi\rangle = \frac{|\tfrac{1}{2},\tfrac{1}{2};\tfrac{1}{2},0;k\rangle + \alpha |\tfrac{1}{2},\tfrac{1}{2};\tfrac{3}{2},1;k\rangle + \beta |\tfrac{1}{2},\tfrac{1}{2};\tfrac{1}{2},1;k\rangle}{\sqrt{1+|\alpha|^2+|\beta|^2}}
\end{eqnarray}
where the intermixing coefficients $\alpha$ and $\beta$ determine the amount of intermixing of the $\Gamma_8^v$ ($J=\tfrac{3}{2}, L_{\text{B}}=1$) and $\Gamma_7^v$ ($J=\tfrac{1}{2}, L_{\text{B}}=1$) bands into electron ground state, which originates predominantly from the $\Gamma_6^c$ ($J=\tfrac{1}{2}, L_{\text{B}}=0$) band. After diagonalizing the Hamiltonian, we find the intermixing coefficients to be:
\begin{eqnarray*}
\alpha &=& i\sqrt{\frac{2}{3}} \frac{\lambda - \frac{\hbar^2 k^2}{2 m_0}}{k P_0} \frac{\left(\gamma_1-2\gamma_{23}\right)\frac{\hbar^2 k^2}{2 m_0} + \left(E_g+\Delta+\lambda\right)}{\left(\gamma_1-2\gamma_{23}\right)\frac{\hbar^2 k^2}{2 m_0} +  \left(E_g+\frac{2}{3}\Delta+\lambda\right)} \\
\beta &=& i\sqrt{\frac{1}{3}} \frac{\lambda - \frac{\hbar^2 k^2}{2 m_0}}{k P_0} \frac{\left(\gamma_1-2\gamma_{23}\right)\frac{\hbar^2 k^2}{2 m_0} + \left(E_g+\lambda\right)}{\left(\gamma_1-2\gamma_{23}\right)\frac{\hbar^2 k^2}{2 m_0} +  \left(E_g+\frac{2}{3}\Delta+\lambda\right)}
\end{eqnarray*}
where $\lambda=\lambda(k)$ is the confinement energy (i.e. the energy of the state above the conduction band edge), given by one of the roots of $|{\cal H}_{\frac{1}{2},+\frac{1}{2}}-\lambda I|=0$.

\subsection{Spheres} \label{sec:spheres}

\begin{figure}
\begin{center}
\includegraphics[width=\columnwidth]{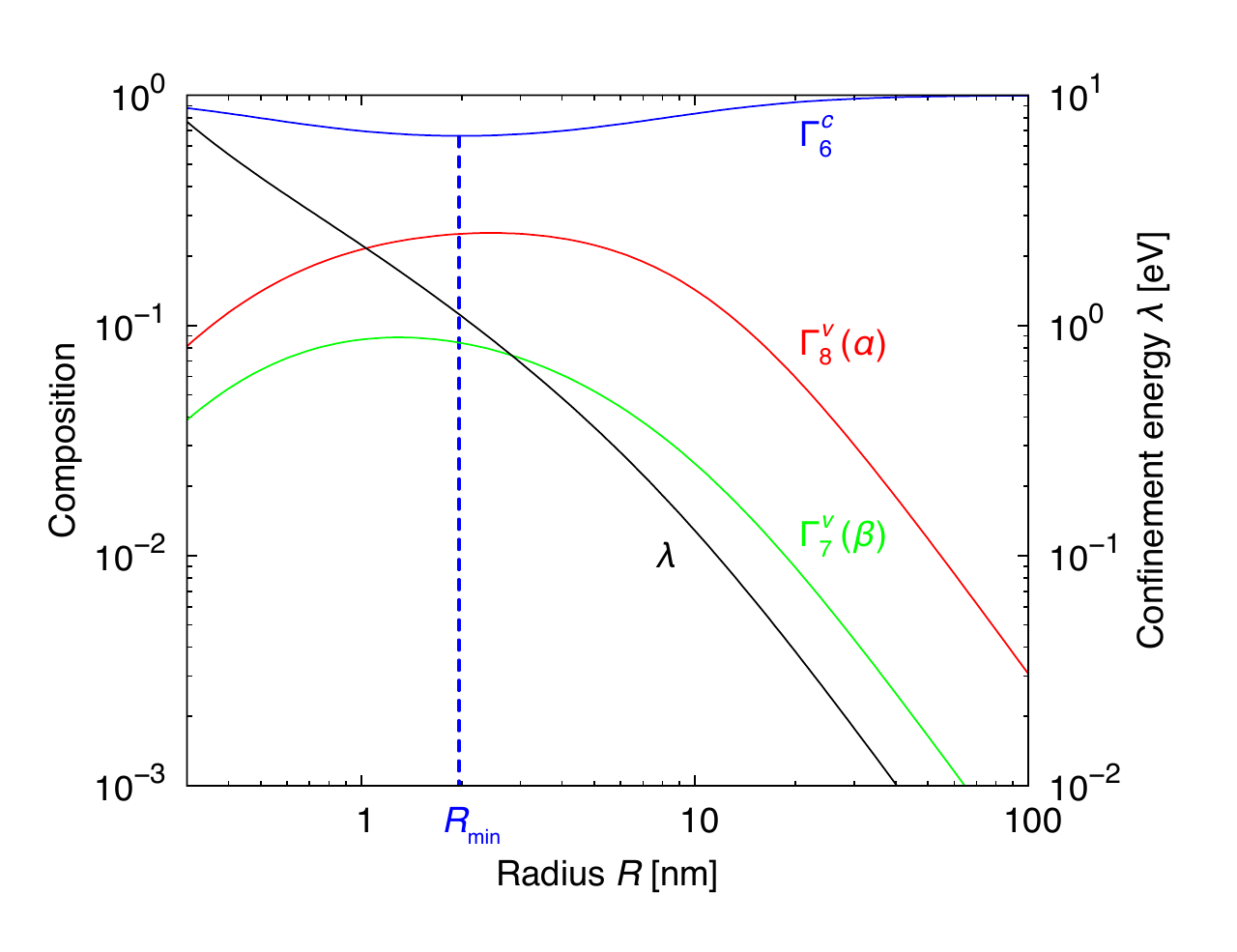}
\caption{Radius dependence of the confinement energy $\lambda$ and composition of the electron ground state of an InAs sphere. The composition is given in terms of the conduction band $\Gamma_6^c$ (blue), and valence bands $\Gamma_8^v$ (red) and $\Gamma_7^v$ (green) contributions.}
\label{fig:mix_spheres}
\end{center}
\end{figure}

We start by examining solid spheres, for which the confining potential is 
\begin{eqnarray}
V(r) = \left\{
\begin{array}{ll}
0 & r \leq R \\
\infty & r > R
\end{array}
\right.
\end{eqnarray}
where $R$ is the radius of the sphere. The wave function needs to be normalizable at the origin of the sphere, hence only spherical Bessel functions $j_l(kr)$ contribute to the envelope wave function ($\xi=0$). We assume that the electron ground state predominantly originates from conduction band states. We therefore choose the approximate boundary condition $\langle r,\theta,\phi|\tfrac{1}{2},+\tfrac{1}{2};\tfrac{1}{2},0;k\rangle |_{r=R}=0$, from which the relation $k=\tfrac{\pi}{R}$ follows.

\begin{figure}
\begin{center}
\includegraphics[width=\columnwidth]{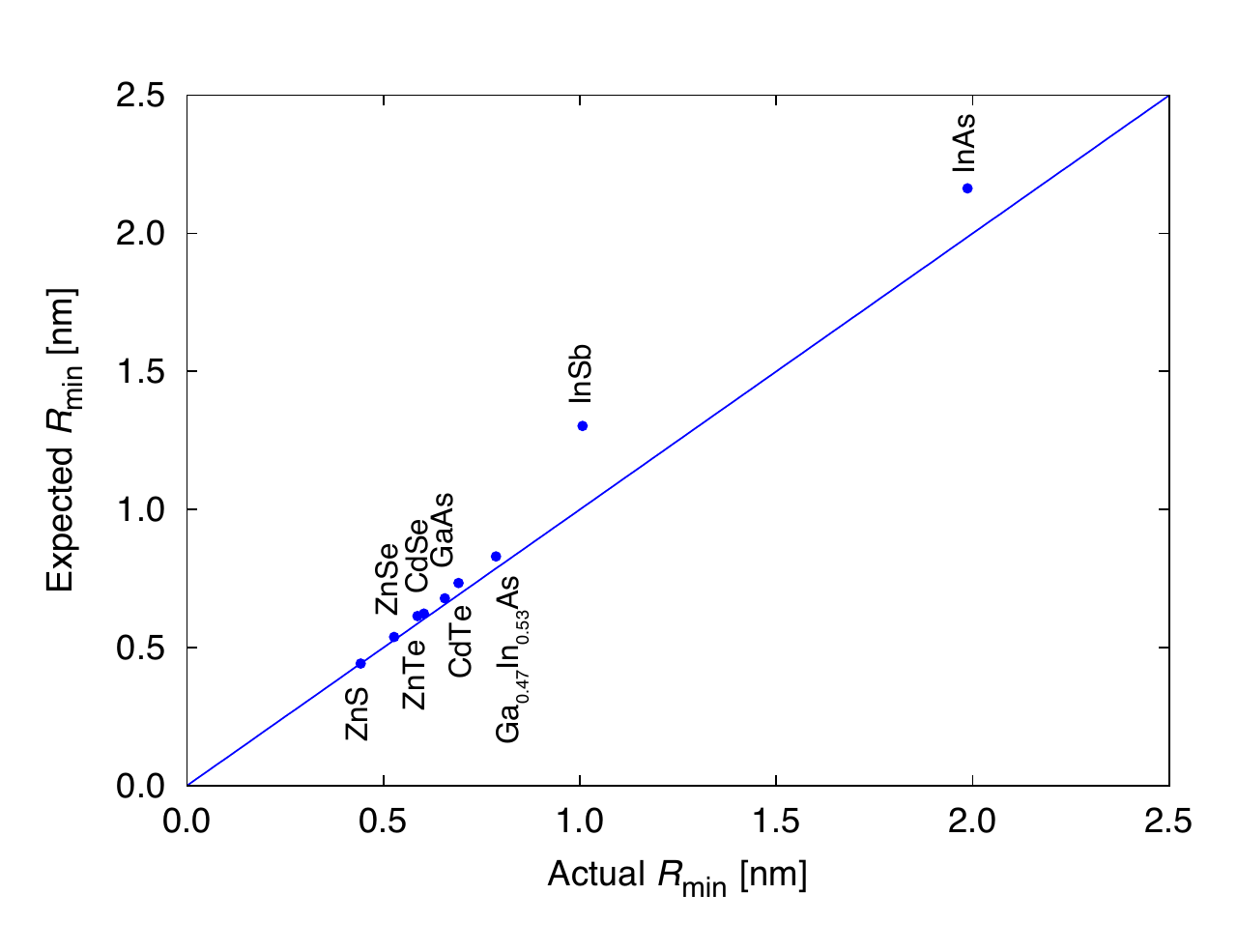}
\caption{The actual and expected (see Eq.~\ref{eq:rmin}) radius $R_{\text{min}}$ where the electron ground state of spheres of different materials have the smallest conduction band contribution.}
\label{fig:mix_spheres_rmin}
\end{center}
\end{figure}

\begin{figure}
\begin{center}
\includegraphics[width=\columnwidth]{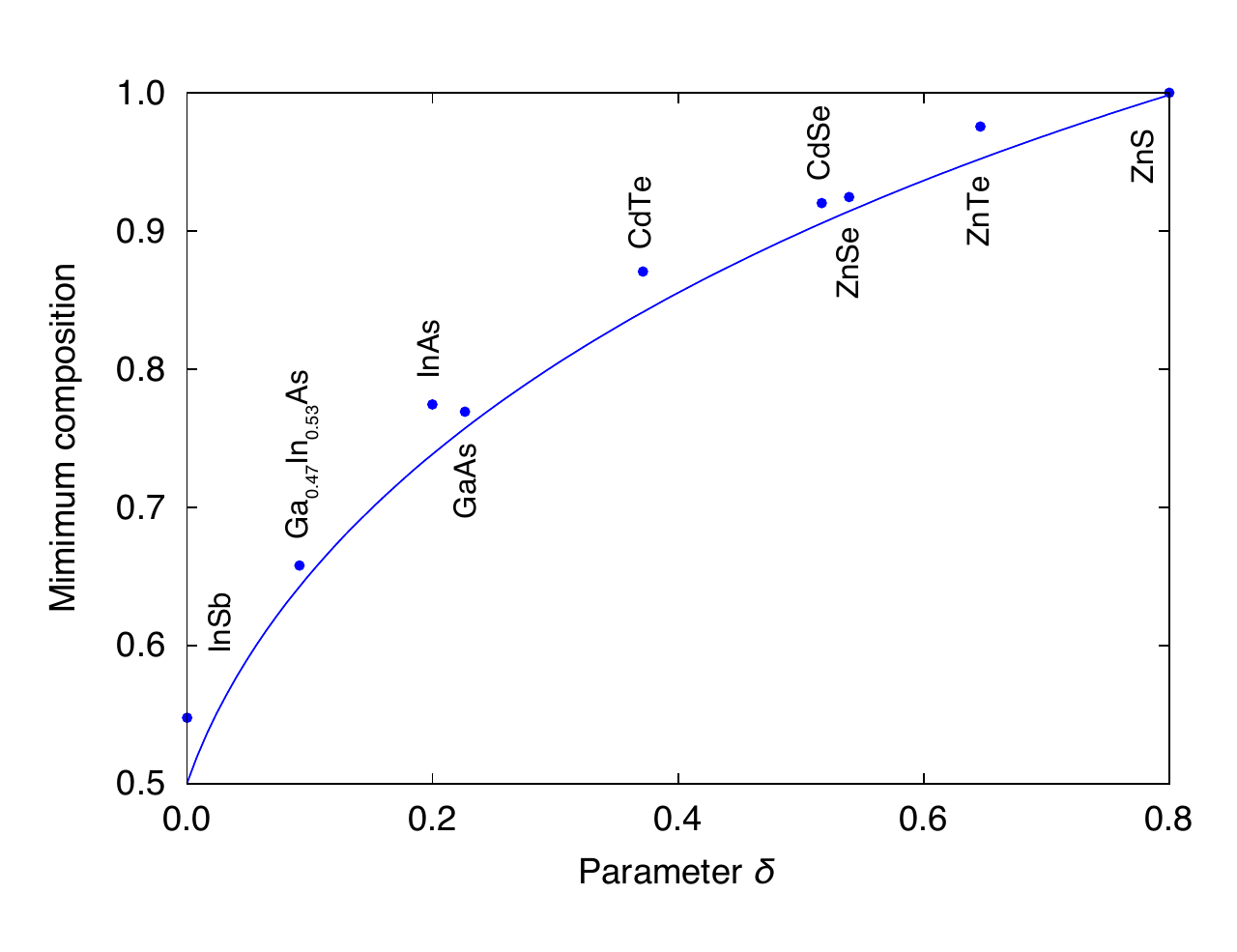}
\caption{The minimum contribution of the conduction band of spheres of different materials as function of the parameter $\delta$, along with Eq.~\ref{eq:mincomp}.}
\label{fig:mix_spheres_mincomp}
\end{center}
\end{figure}

In Fig.~\ref{fig:mix_spheres} we plot the radius dependence of the confinement energy and composition of the electron ground state for an InAs sphere. The intermixing of the valence bands is never very large ($<30$~\%), so that their influence can be regarded as a perturbation on the predominantly conduction band-like state. This perturbation is proportional to the ratio of the coupling of the bands and the energetic splitting between them. The former is constant in our problem ($\sim k P_0$), but the latter is not and leads to the maximum around $2$~nm. At large $R$ (small $k$) the energetic splitting is given mainly by the energy differences between the bands ($E_g$ for $\alpha$, $E_g+\Delta$ for $\beta$), resulting in a $1/R$ dependence of the intermixing. At small $R$ (large $k$) the energetic splitting is dominated by the free kinetic energy of the conduction and valence bands, which results in a $R$-dependence of the intermixing. The intermixing therefore peaks when the free kinetic energy is equal to $E_g$ for $\alpha$, or $E_g+\Delta$ for $\beta$. This condition can be expressed analytically in the limit of zero spin-orbit coupling, when the free kinetic energy of the valence band can be expressed in a simple manner:
\begin{eqnarray}
\frac{\hbar^2 k^2}{2 m_0} + (\gamma_1+4\gamma_{23})\frac{\hbar^2 k^2}{2 m_0} = E_g
\end{eqnarray}
From this condition we can extract the radius $R_{\text{min}}$ at which the conduction band has the smallest contribution:
\begin{eqnarray}
R_{\text{min}} = \pi\hbar\sqrt{\frac{1+\gamma_1+4\gamma_{23}}{2 m_0 E_g}} \label{eq:rmin}
\end{eqnarray}
The minimum radius depends therefore on the effective hole mass and band gap energy, which we exemplified by showing $R_{\text{min}}$ in Fig.~\ref{fig:mix_spheres_rmin} for various semiconductor materials. Alongside the actual $R_{\text{min}}$, we also plot the expected $R_{\text{min}}$ on basis of the above formula. It can be seen that the above formula is a good predictor for $R_{\text{min}}$, as long as $\Delta/E_g\ll1$ (hence not for InSb and InAs). We find that the contribution of the conduction band at $R_{\text{min}}$ can be expressed as:
\begin{eqnarray}
\text{Min. comp.} = \frac{1}{2+\delta-\sqrt{\delta(\delta+2)}} \approx \frac{1}{2} + \frac{1}{2\sqrt{2}}\sqrt{\delta}\nonumber\label{eq:mincomp}\\
\end{eqnarray} 
where
\begin{eqnarray}
\delta = (1+\gamma_1+4\gamma_{23})\frac{\hbar^2 E_g}{m_0 P_0^2}
\end{eqnarray}
In Fig.~\ref{fig:mix_spheres_mincomp} we plot for various materials the actual minimum contribution and the expected contribution based on the above formula. Since we assumed $\Delta=0$, the formula is overestimating the intermixing of the valence band and can be regarded as an lower limit of the actual minimum contribution. It can be seen that the minimum contribution is always more than 50\%, and increases with $E_g$ and a smaller effective hole mass, which explains why In-compounds have a stronger valence band mixing than Zn-compounds.

\begin{figure}
\begin{center}
\includegraphics[width=\columnwidth]{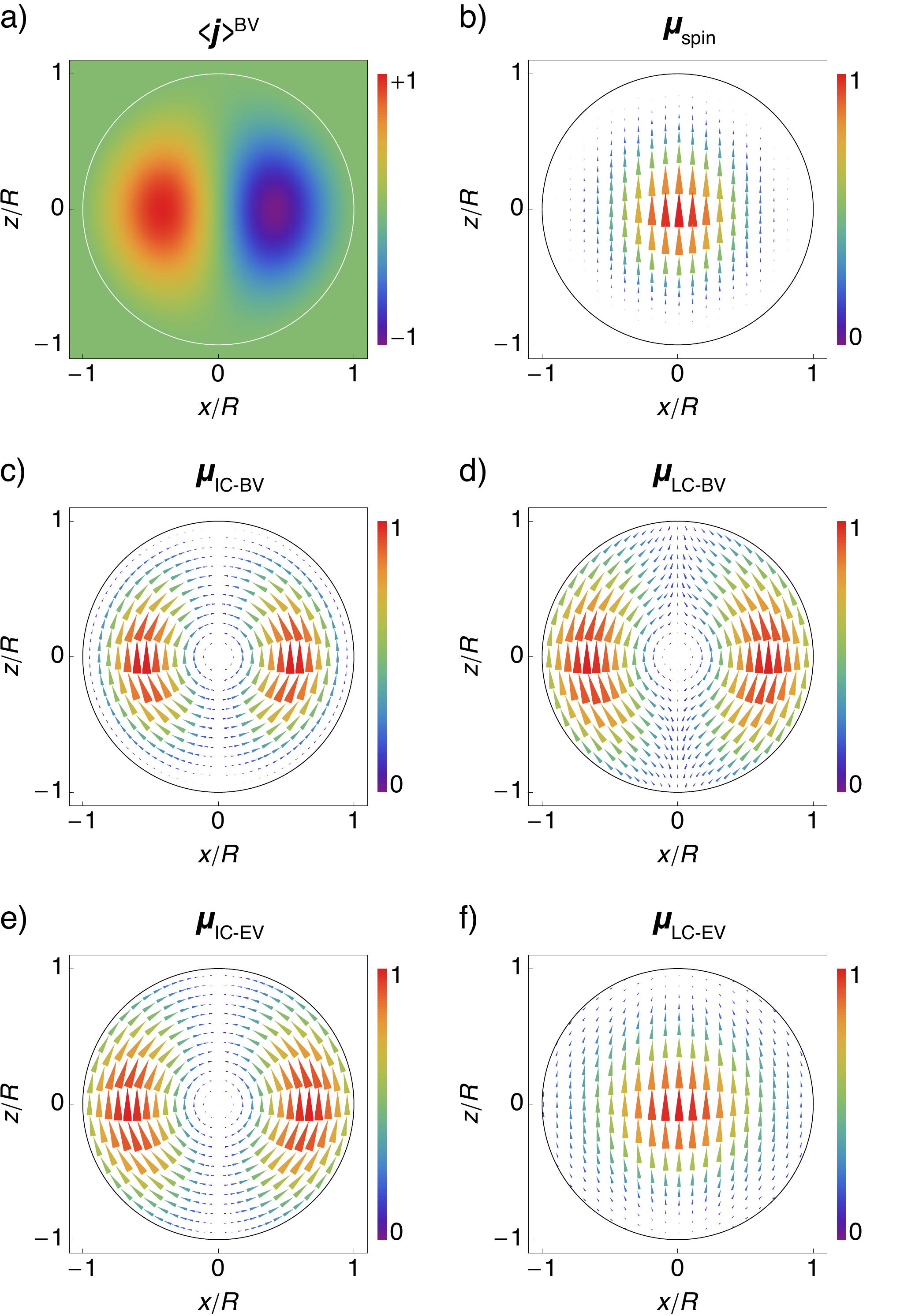}
\caption{(a) The spatial distribution of the normalized magnitude of the ${\bf e}_{y}$-component of $\langle {\bf j}\rangle^{\text{BV}}$ of a sphere. This current density peaks roughly at $R/2$ and resembles a current loop. (b-f) The magnetic moment density of the different components contributing to the orbital moment (c-f) and the spin moment (b) of a sphere. It can clearly be observed that  ${\boldsymbol \mu}_{\text{spin}}$ and ${\boldsymbol \mu}_{\text{LC-EV}}$ have an even spatial symmetry, whereas the other orbital moments have an odd spatial symmetry. All figures are $xz$-cross-sections, the white/black circles mark the boundary of the sphere.}
\label{fig:mu_spatial_spheres}
\end{center}
\end{figure}

Now that the wave function of the ground state is known, we can proceed by calculating the itinerant BV related current density $\langle {\bf j}\rangle^{\text{BV}}$:
\begin{eqnarray}
\langle{\bf j}\rangle^{\text{BV}} = -\frac{e P_0}{2\sqrt{6} \pi \hbar}\frac{\text{Im}\big\{\alpha-\sqrt{2}\beta\big\}}{1+|\alpha|^2+|\beta|^2} j_0(kr)j_1(kr) \sin(\theta) {\bf e}_{\phi}\nonumber\\
\end{eqnarray}
As anticipated in Sec.~\ref{sec:orbmu}, the spatial distribution of this current is governed by the product of the envelope wave functions associated with the conduction band, $j_0(kr)$, and the valence band, $j_1(kr)$. It therefore resembles a current loop extended throughout the quantum dot and peaks at about $R/2$, see Fig.~\ref{fig:mu_spatial_spheres}(a). Note that this current is proportional to the factor $\text{Im}\big\{\alpha-\sqrt{2}\beta\big\}$, which can be expressed as:
\begin{eqnarray}
&& \text{Im}\big\{\alpha-\sqrt{2}\beta\big\} = \\
&& \hspace{6mm} \sqrt{\frac{2}{3}} \frac{\lambda - \frac{\hbar^2 k^2}{2 m_0}}{k P_0} \frac{\Delta}{\left(\gamma_1-2\gamma_{23}\right)\frac{\hbar^2 k^2}{2 m_0} +  \left(E_g+\frac{2}{3}\Delta+\lambda\right)}\nonumber
\end{eqnarray}
showing explicitly the spin-orbit correlated nature of this current: it directly depends on the spin-orbit coupling $\Delta$. It proves interesting to trace the exact origin of this current. The direction of $\langle{\bf j}\rangle^{\text{BV}}$ comes from the momentum matrix elements $\langle u_i | \nabla | u_j \rangle$, which are only non-zero if $i$ labels a conduction band state and $j$ a valence band state. Because the divergence of $\langle{\bf j}\rangle^{\text{BV}}$ must be zero and the spherical symmetry of the quantum dot, the current has to flow in the ${\bf e}_{\phi}$-direction. The matrix elements associated with this direction can be written as $\langle u_i | \tfrac{1}{r \sin \theta}\tfrac{i}{\hbar}L_{\text{B},z} | u_j\rangle {\bf e}_{\phi}$, hence only Bloch states with non-zero $L_{\text{B},z}$ will contribute to $\langle{\bf j}\rangle^{\text{BV}}$. Only three of such states are present in the $|F,F_z\rangle=|\tfrac{1}{2},+\tfrac{1}{2}\rangle$ electron ground state\cite{Sercel1990}: $|J,J_z;L_{\text{B}},L_{\text{B},z}\rangle=|\tfrac{3}{2},+\tfrac{3}{2};1,+1\rangle$, $|\tfrac{3}{2},-\tfrac{1}{2};1,-1\rangle$, and $|\tfrac{1}{2},-\tfrac{1}{2};1,-1\rangle$. The former will create a current opposite to the latter two due to the different orientation of $L_{\text{B}}$. The degree of cancellation depends on the strength of the spin-orbit coupling, as this will tune the presence of the $|\tfrac{1}{2},-\tfrac{1}{2};1,-1\rangle$ (split-off) state. This mechanism has also been identified to determine the bulk $g$-factor of semiconductors\cite{Yafet1963}.

\begin{figure}[t]
\begin{center}
\includegraphics[width=\columnwidth]{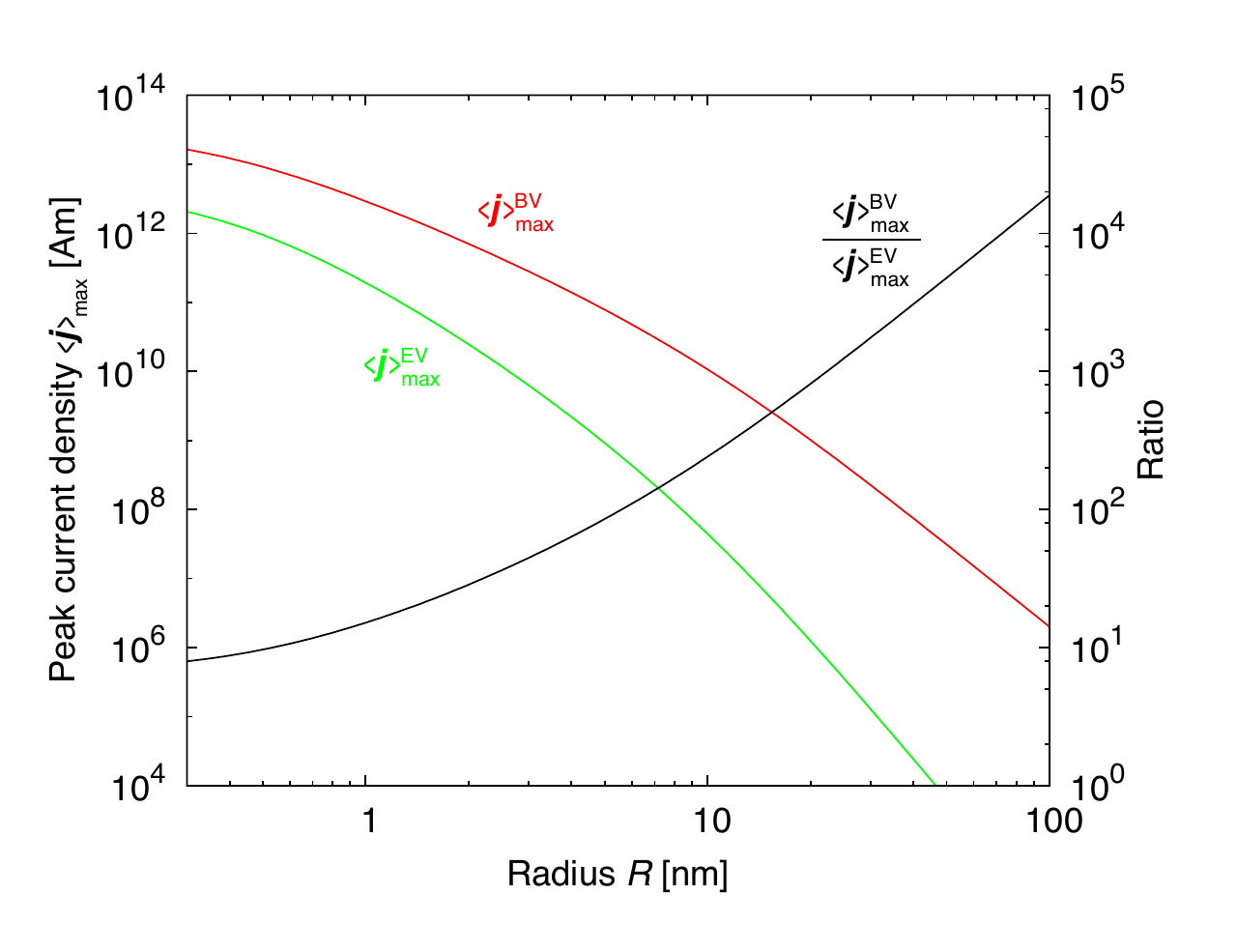}
\caption{The radius dependence of the peak current densities $\langle {\bf j} \rangle^{\text{BV}}$ (red) and $\langle {\bf j} \rangle^{\text{EV}}$ (green), and their ratio (black) of an InAs sphere. The Bloch velocity related current is $\geq 10$ times the envelope velocity related current for radii $R\geq1$~nm.}
\label{fig:j_spheres}
\end{center}
\end{figure}

The itinerant EV related current density $\langle {\bf j}\rangle^{\text{EV}}$ can be more generally calculated using the general envelope state $|k,L_{\text{E}},L_{\text{E},z}\rangle$:
\begin{eqnarray}
\langle {\bf j} \rangle^{\text{EV}}_{|k,L_{\text{E}},L_{\text{E},z}\rangle} = \frac{2 e \hbar}{m_0} L_{\text{E},z} \frac{j_{L_{\text{E},z}}(kr)^2}{\pi r \sin \theta} |Y_{L_{\text{E}}}^{L_{\text{E},z}}(\theta,\phi)|^2 {\bf e}_{\phi}\nonumber\\
\end{eqnarray}
by which $\langle {\bf j}\rangle^{\text{EV}}$ of the electron ground state becomes:
\begin{eqnarray}
\langle {\bf j}\rangle^{\text{EV}} = - \frac{e \hbar}{8 \pi m_0} \frac{|\alpha|^2-2|\beta|^2}{1+|\alpha|^2+|\beta|^2} \frac{j_1(kr)^2}{r} \sin(\theta){\bf e}_{\phi}
\end{eqnarray}
The envelope wave function associated with the conduction band has $L_{\text{E}}=0$ and therefore does not contribute to $\langle {\bf j}\rangle^{\text{EV}}$; this current originates solely from the valence band. The spatial distribution is therefore governed by the square of the valence band envelope wave functions, i.e. $j_1(kr)^2$, though it has the same spatial symmetry as $\langle {\bf j}\rangle^{\text{BV}}$. We again emphasize that this current has as spin-orbit correlated nature: the factor $|\alpha|^2-2|\beta|^2$ is directly proportional to $\Delta$. We plot both the peak current densities $\langle {\bf j}\rangle^{\text{BV}}_{\text{max}}$ and $\langle {\bf j}\rangle^{\text{EV}}_{\text{max}}$ in Fig.~\ref{fig:j_spheres}, together with the ratio between them. It can be clearly observed that the Bloch velocity related current is $\geq 10$ times larger than the envelope velocity related current for realistic sizes, as was anticipated in Sec.~\ref{sec:orbmu}.

Using Eqs.~\ref{eq:mu_ICBV}~and~\ref{eq:mu_LCBV} we can plot the orbital moment densities related to the Bloch velocity, see Fig.~\ref{fig:mu_spatial_spheres}(c)~and~(d). As expected, their spatial distributions have the same (odd) spatial symmetry, though they differ slightly in the exact distribution. We can do the same for the envelope velocity related orbital momenta in Fig.~\ref{fig:mu_spatial_spheres}(e)~and(f), using Eqs.~\ref{eq:mu_ICEV}~and~\ref{eq:mu_LCEV}. As expected, the spatial distribution of ${\boldsymbol \mu}_{\text{IC-EV}}$ has an odd spatial symmetry, whereas ${\boldsymbol \mu}_{\text{LC-EV}}$ has an even spatial symmetry. The latter shares this symmetry with the spin moment density, which is plotted in Fig.~\ref{fig:mu_spatial_spheres}(b) using Eq.~\ref{eq:muspin}. As discussed in Ref.~\onlinecite{Bree2014}, these different symmetries can have substantial consequences, for example for the hyperfine coupling or interactions with nearby magnetic moments. Although the spin moment density seems to be parallel to the $z$-direction, there is in fact a very small $x$-component due to intermixing of the valence band states. This component has an odd spatial symmetry and is so small, $\leq 0.1\%$ of the $z$-component, that we have neglected it for the plot.

\begin{figure}
\begin{center}
\includegraphics[width=\columnwidth]{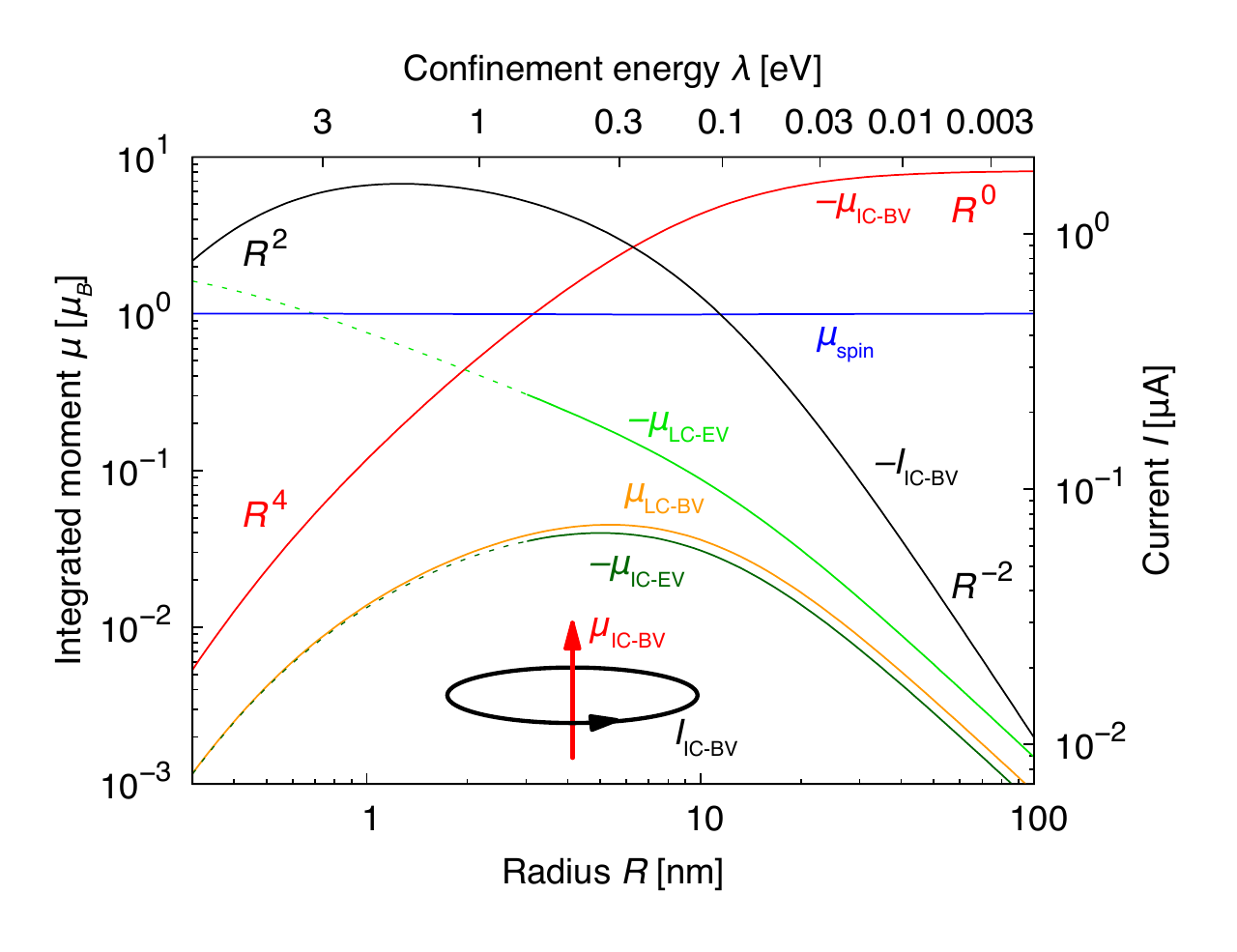}
\caption{The radius dependence of the different integrated orbital and spin moments and current $I_{\text{IC-BV}}$ of an InAs sphere.}
\label{fig:mu_spheres}
\end{center}
\end{figure}

By integrating the moment densities over the whole state, we can compute the different contributions to the integrated orbital magnetic moment:
\begin{eqnarray}
{{\boldsymbol \mu}_{\text{IC-BV}}} &=& -{\mu_B}\sqrt{\frac{2}{3}}\frac{m_0 P_0 R}{\pi\hbar^2} \frac{\text{Im}\{\alpha - \sqrt{2}\beta\}}{1+|\alpha|^2+|\beta|^2} {\bf e}_z \\
{{\boldsymbol \mu}_{\text{LC-BV}}} &=& +{\mu_B}\left[ \frac{1}{3}\frac{|\alpha|^2 - 2 |\beta|^2}{1+|\alpha|^2+|\beta|^2} + \frac{2}{9}\frac{\text{Im}\{\alpha-\sqrt{2}\beta\}^2}{1+|\alpha|^2+|\beta|^2}\right] {\bf e}_z \nonumber \\
\\
{{\boldsymbol \mu}_{\text{IC-EV}}} &=& -{\mu_B} \frac{1}{3}\frac{|\alpha|^2-2|\beta|^2}{1+|\alpha|^2+|\beta|^2} {\bf e}_z \\
{{\boldsymbol \mu}_{\text{LC-EV}}} &=& -{\mu_B}\frac{2}{3}\sqrt{\frac{2}{3}} \frac{\pi P_0}{R} \frac{\text{Im}\{\frac{1}{E_g}\alpha-\frac{\sqrt{2}}{E_g+\Delta}\beta\}}{1+|\alpha|^2+|\beta|^2} {\bf e}_z
\end{eqnarray}
and likewise we can calculate the integrated spin moment:
\begin{eqnarray}
{{\boldsymbol \mu}_{\text{spin}}} = {\mu_B}\left[1-\left(\frac{2}{3}\right)^2\frac{\text{Im}\{\alpha-\sqrt{2}\beta\}^2}{1+|\alpha|^2+|\beta|^2}\right] {\bf e}_z
\end{eqnarray}
In Fig.~\ref{fig:mu_spheres} we plot these moments as function of radius $R$ for an InAs sphere. For a wide range of radii, the dominant contribution to the orbital moment is ${\boldsymbol \mu}_{\text{IC-BV}}$. This was to be expected: the largest moment is generated when both the lever arm (itinerant current) and momentum (Bloch velocity) are largest. We will therefore first concentrate on ${\boldsymbol \mu}_{\text{IC-BV}}$. In the limit of infinite radius $R$ (i.e. the bulk limit), ${\boldsymbol \mu}_{\text{IC-BV}}$ reduces to the Roth formula\cite{Roth1959}:
\begin{eqnarray}
\lim_{R\rightarrow\infty}{{\boldsymbol \mu}_{\text{IC-BV}}} = - {\mu_B}\frac{\Delta}{3E_g(E_g+\Delta)} \frac{2 m_0 P_0^2}{\hbar^2} {\bf e}_z = {{\boldsymbol \mu}_{\text{Roth}}}\nonumber\\
\end{eqnarray}
As the radius becomes smaller, ${\boldsymbol \mu}_{\text{IC-BV}}$ quenches since the orbital extend (the lever arm) of the envelope wave function becomes smaller. The current distribution associated with ${\boldsymbol \mu}_{\text{IC-BV}}$ resembles a current loop, as can be seen in Fig.~\ref{fig:mu_spatial_spheres}(a). It proves insightful to make an analogy with a simple current loop, carrying a current $I$ at radius $R$, generating a moment:
\begin{eqnarray}
\mu_{\text{loop}} = \pi I R^2
\end{eqnarray}
This immediately shows that there should be a $R^2$-dependence on the orbital moment. We can formally verify this dependency by calculating the current $I_{\text{IC-BV}}$ in the spheres:
\begin{eqnarray}
I_{\text{IC-BV}} &=& \int \langle {\bf j} \rangle^{\text{BV}} \cdot {\bf n}~da \\
&=& -\frac{e P_0}{\sqrt{6}\pi \hbar R} \frac{\text{Im}\left\{\alpha-\sqrt{2}\beta\right\}}{1+|\alpha|^2+|\beta|^2}  \int_0^{2 \pi} \frac{\sin\chi}{\chi}~d\chi
\end{eqnarray}
which is also plotted in Fig.~\ref{fig:mu_spheres}. It can be immediately verified that the analogy with the classical current loop holds: the ratio between the current $I_{\text{IC-BV}}$ and orbital moment ${\boldsymbol \mu}_{\text{IC-BV}}$ has a $R^2$-dependence. The mechanism leading to quenching of ${\boldsymbol \mu}_{\text{IC-BV}}$ is therefore an interplay of two effects: quantum confinement limits the extension of the envelope wave function and reduces thereby the lever arm, while intermixing of the valence bands determines the amount of current that circulates in the sphere.

\begin{figure}
\begin{center}
\includegraphics[width=\columnwidth]{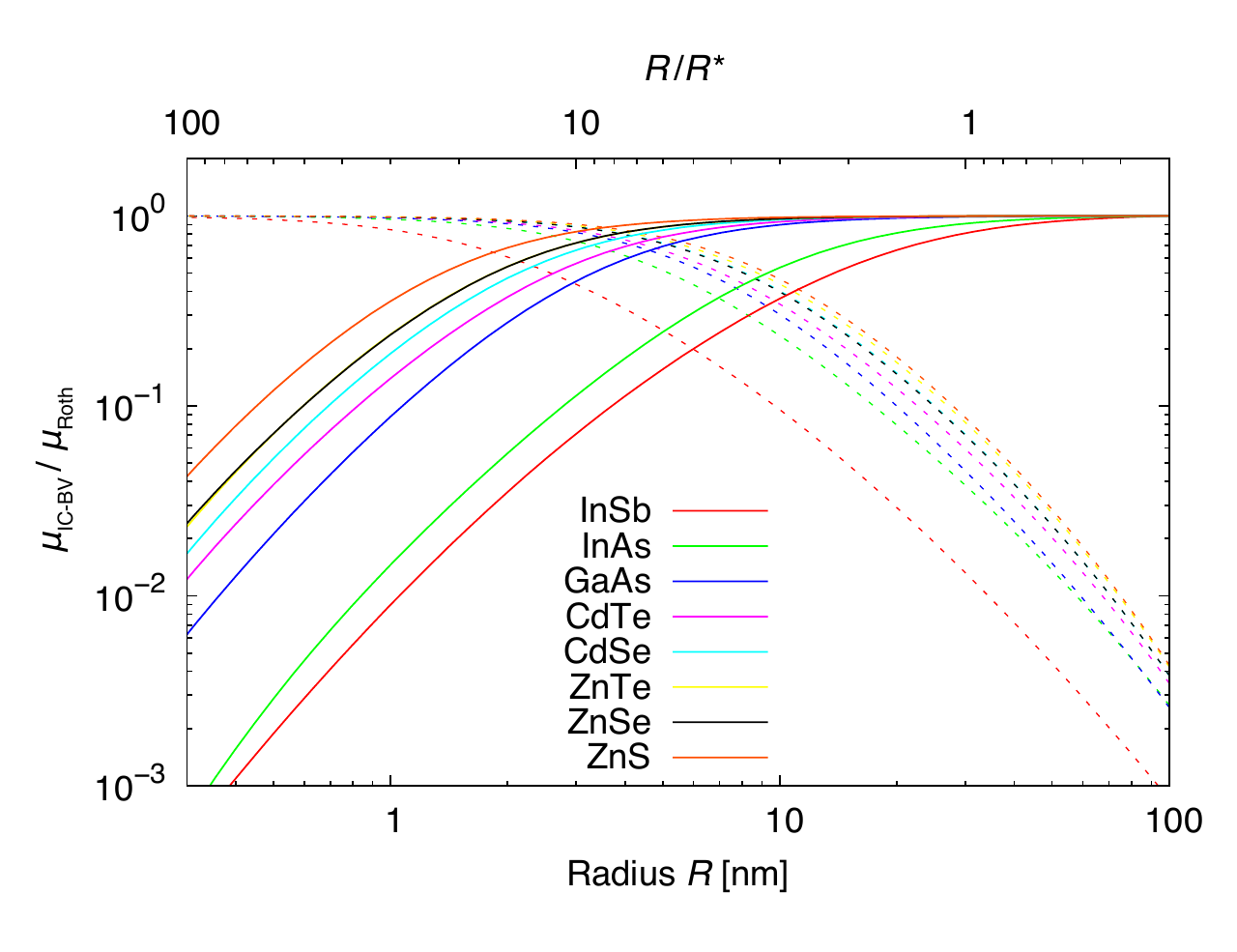}
\caption{The radius dependence of ${\boldsymbol \mu}_{\text{IC-BV}}/{\boldsymbol \mu}_{\text{Roth}}$ of spheres of various semiconducting materials (continuous lines). The same quantity is also plotted against the normalized radius $R/R^*$ (dotted lines).}
\label{fig:mu_spheres_materials}
\end{center}
\end{figure}

The other contributions to the orbital moment have a non-monotonic dependence on $R$. As expected, these contributions to the orbital moment are small compared to ${\boldsymbol \mu}_{\text{IC-BV}}$, since either the lever arm (localized currents), or the momentum (envelope velocity) is small. In particular, ${\boldsymbol \mu}_{\text{LC-BV}}$ is small since $\langle u_i | {\bf L}_{\text{B}} | u_j \rangle$ is only non-zero for Bloch functions not involving the conduction band. Therefore ${\boldsymbol \mu}_{\text{LC-BV}}$ is (more or less) proportional to the intermixing of valence bands and is always small, which can be verified by comparison of Figs.~\ref{fig:mix_spheres}~and~\ref{fig:mu_spheres}. A similar argument holds for ${\boldsymbol \mu}_{\text{IC-EV}}$, which originates from $\langle {\bf j} \rangle^{\text{EV}}$ and is therefore directly proportional to the amount of the intermixing of valence states, since the conduction band envelope has $L_{\text{E}}=0$. Note that ${\boldsymbol \mu}_{\text{LC-BV}}\approx-{\boldsymbol \mu}_{\text{IC-EV}}$, so these moments cancel each other when added to the total orbital moment. This (near) cancellation arises from the fact that $L_{\text{B},z} = -L_{\text{E},z}$ for most bands contributing to the electron ground state. A more detailed analysis of this effect will be performed for the disks with hard-wall boundaries at the end of Sec.~\ref{sec:harddisks}. Lastly, the behavior of ${\boldsymbol \mu}_{\text{LC-EV}}$ stands out: it gets larger for smaller $R$. As can be seen from Eq.~\ref{eq:LCEV}, ${\boldsymbol \mu}_{\text{LC-EV}}$ is proportional to $\nabla F_j({\bf r}_s)$ and will therefore become larger as the quantum dot becomes smaller. The envelope function approximation becomes less accurate as $R$ becomes smaller, and quantities involving the gradient of the envelope wave function will be affect first. We therefore plot the moments related to the envelope velocity dotted for $R\leq3$~nm.

The spin moment is almost constant at one $\mu_B$, dropping about 1\% at a radius of $7$~nm. Even though a sizable amount of valence states mix into the electron ground state, the effect on the spin moment is negligible due to the same cancellation mechanism discussed for $\langle{\bf j}\rangle^{\text{BV}}$. In fact, the deviation of the spin moment is proportional to (square of) the same factor $\text{Im}\{\alpha -\sqrt{2}\beta\}$. This means that these deviations vanish in absence of spin-orbit coupling.

\begin{figure}
\begin{center}
\includegraphics[width=\columnwidth]{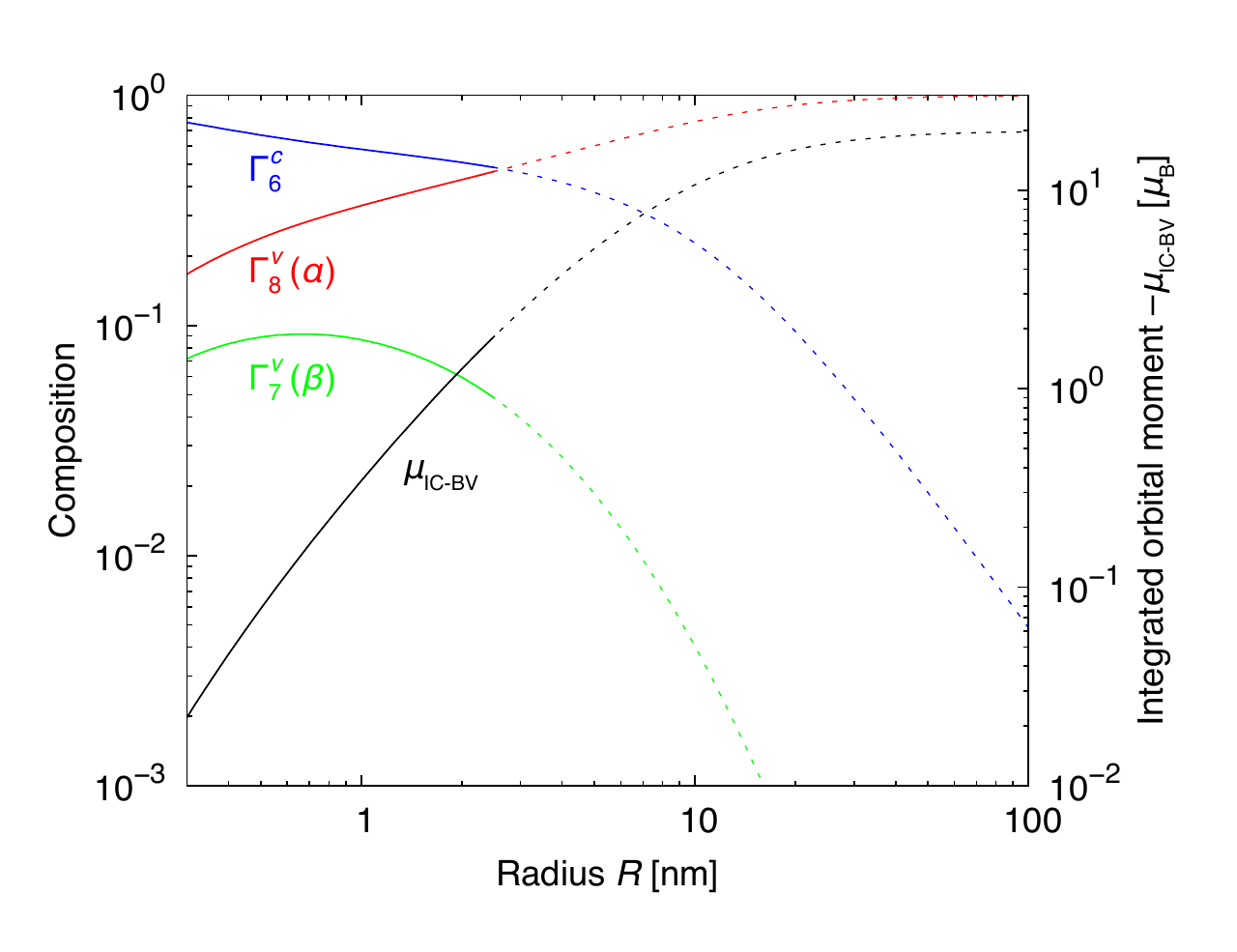}
\caption{The radius dependence of the composition and integrated orbital moment ${\boldsymbol \mu}_{\text{IC-BV}}$ of a HgTe sphere. The calculation is only valid for $R<2.7$~nm, hence the lines are dotted for $R>2.7$~nm.}
\label{fig:mu_spheres_HgTe}
\end{center}
\end{figure}

Up to now we have used InAs as the constituent material of the spheres, but it is also interesting to see how the orbital moment is quenched in other semiconductors. We therefore show in Fig.~\ref{fig:mu_spheres_materials} the dominant orbital moment ${\boldsymbol \mu}_{\text{IC-BV}}$ normalized to the Roth formula\cite{Roth1959} for spheres of various materials. Both the current distribution and the mechanism leading to quenching of the orbital moment are similar to what was found for InAs spheres. The onset of quenching of the orbital moment, however, depends on the material of the sphere. This observation can be made more explicit by analyzing ${\boldsymbol \mu}_{\text{IC-BV}}$ in the limit of small $R$:
\begin{eqnarray}
\lim_{R \rightarrow 0} \frac{{\boldsymbol \mu}_{\text{IC-BV}}}{{\boldsymbol \mu}_{\text{Roth}}} &=& \frac{E_g(E_g+\Delta)}{(1+\gamma_1-2\gamma_{23})(1+\gamma_1+4\gamma_{23})} \left[\frac{2m_0R^2}{\hbar^2\pi^2}\right]^2 \nonumber\\
&\equiv& \left(\frac{R}{R^*}\right)^4
\end{eqnarray}
where we have defined a material-dependent radius $R^*$, which renormalizes ${\boldsymbol \mu}_{\text{IC-BV}}$ at small $R$ (see dotted lines in Fig.~\ref{fig:mu_spheres_materials}). A large $R^*$ means that the quenching starts at relatively large $R$, and arises from either a small effective hole mass, a small band gap, or a small spin-orbit coupling. This explains why spheres from In-compounds start to quench at larger $R$ compared to spheres from Zn-compounds.

Besides semiconductors, it is also interesting to see the effects of spin-orbit correlated currents in semimetals. We focus here on zinc-blende HgTe, of which the synthesis of small colloidal quantum dots is well established\cite{Keuleyan2011}. Compared to the previously studied materials, the ordering of the bands at the $\Gamma$-point is different in HgTe: the $\Gamma_6^c$-band has a lower energy than the $\Gamma_{7,8}^v$-bands\cite{Groves1967}. However, the first empty band (i.e. the $\Gamma_8^v$-band) is connected to both the $X_6^c$-point and $L_6^c$-point, meaning that the character of the band changes at finite $k$ (see Ref.~\onlinecite{Madelung2004}). Consequently, for sufficiently small spheres (large $k$), the electron ground state must have predominantly a $\Gamma_6^c$ character, which can indeed be observed in Fig.~\ref{fig:mu_spheres_HgTe}. Our approach to calculate the electron state assumes that the state is mainly stemming from the $\Gamma_6^c$-band. This assumption is therefore only valid for $R<2.7$~nm, and hence we plot in Fig.~\ref{fig:mu_spheres_HgTe} the curves dotted for $R>2.7$~nm. For small $R$, the current distribution is the same as for the previously studied materials and the integrated orbital moment is quenched in a similar fashion. This demonstrates the general applicability of our approach to calculate spin-orbit correlated currents in nanostructures. For large $R$, the electron ground state is contained in the $F=\tfrac{3}{2}$ subspace, which falls outside the scope of this article. We emphasize that we therefore cannot correctly reproduce the bulk orbital moment (see Table~\ref{table:materials}).

\subsection{Spherical shells} \label{sec:shells}

\begin{figure}
\begin{center}
\includegraphics[width=\columnwidth]{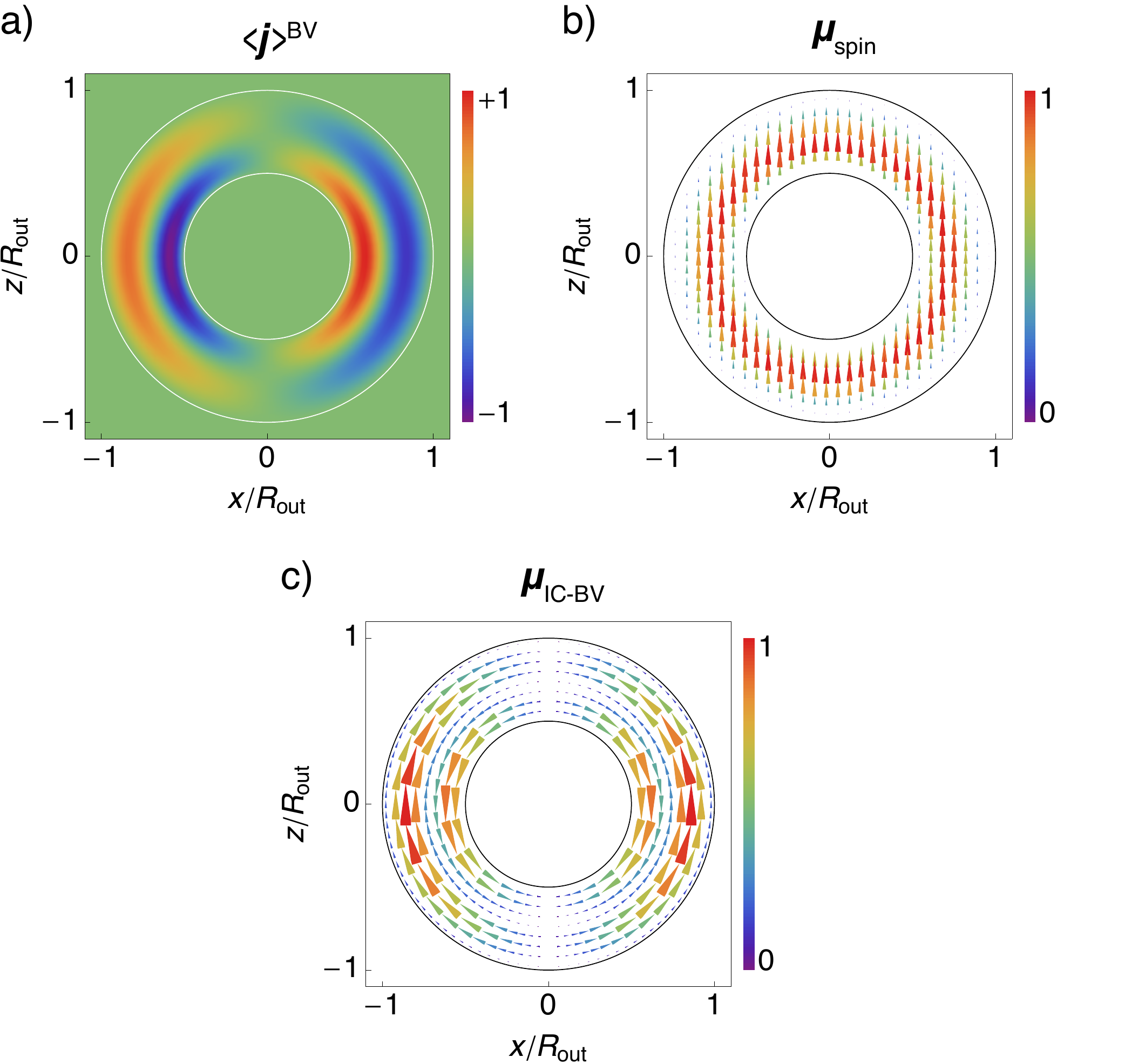}
\caption{(a) The spatial distribution of the normalized magnitude of the ${\bf e}_{y}$-component of $\langle {\bf j}\rangle^{\text{BV}}$ of a spherical shell. Due to the inner surface, an additional current loop is created, circulating oppositely to the outer current loop. (b-c) The magnetic moment density of the most dominant orbital moment (c) and the spin moment (b) of a spherical shell. It can clearly be observed that  ${\boldsymbol \mu}_{\text{spin}}$ has an even spatial symmetry, whereas ${\boldsymbol \mu}_{\text{IC-BV}}$ has an odd spatial symmetry. All figures are $xz$-cross-sections, the white/black circles mark the boundaries of the spherical shell, we choose $R_{\text{in}}/R_{\text{out}}=\tfrac{1}{2}$. Similar to the spheres, we have neglected the very small $x$-component of the spin moment density, see the discussion in Sec.~\ref{sec:spheres}.}
\label{fig:mu_spatial_shell}
\end{center}
\end{figure}

\begin{figure}
\begin{center}
\includegraphics[width=\columnwidth]{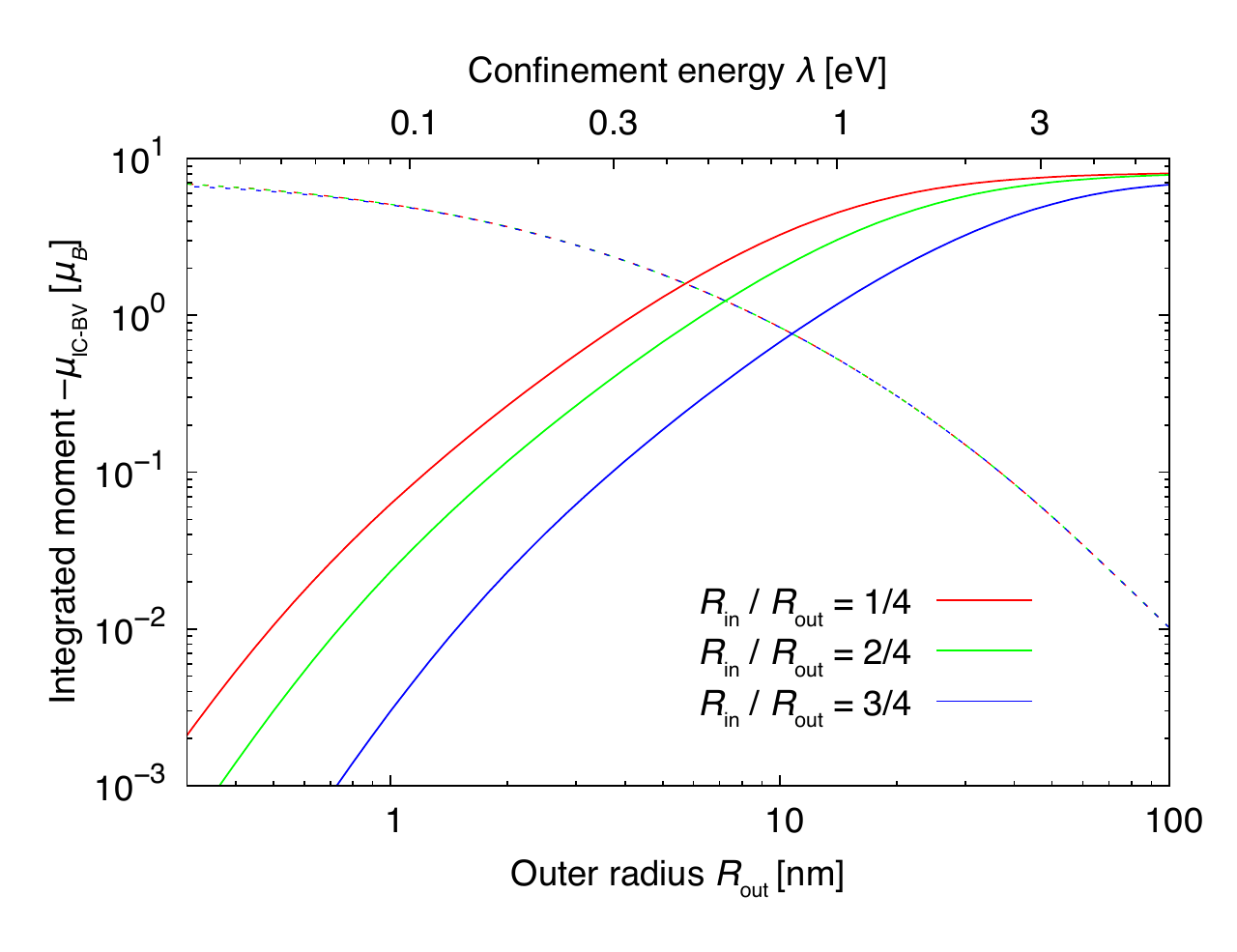}
\caption{The integrated orbital moment ${\boldsymbol \mu}_{\text{IC-BV}}$ of InAs shells for various ratios $R_{\text{in}}/R_{\text{out}}$ as function of the outer radius $R_{\text{out}}$ (continuous lines) and confinement energy $\lambda$ (dotted lines).}
\label{fig:mu_shell}
\end{center}
\end{figure}

In the previous subsection we found that the dominant orbital current is zero at the center of the sphere and peaks at roughly at $R/2$. This suggests that removing material from the center of the sphere would not affect the current distribution in a significant way. We therefore investigate now a spherical shell, where material is indeed removed from the center. The spherical shell is also interesting from another perspective: we found that the currents and moments of the spheres are governed by a single geometrical parameter, the radius $R$ of the sphere. A spherical shell, however, has in principle two independent geometrical parameters: its inner radius $R_{\text{in}}$ and outer radius $R_{\text{out}}$. The confining potential of such a spherical shell is given by:
\begin{eqnarray}
V(r) = \left\{
\begin{array}{lc}
0      & R_{\text{in}} \leq r \leq R_{\text{out}} \\
\infty & \text{elsewhere}
\end{array}
\right.
\end{eqnarray}
The spherical Neumann functions $y_l(r)$ do play a role now, since the origin is not involved in the wave function. Therefore $\xi$ is non-zero and should follow from the boundary conditions. Since the electron ground state predominantly originates from conduction band states, we use the approximate boundary conditions $\langle r,\theta,\phi|\tfrac{1}{2},+\tfrac{1}{2};\tfrac{1}{2},0;k\rangle |_{r=R_{\text{in}},r=R_{\text{out}}}=0$, which leads to the relations:
\begin{eqnarray}
k &=& \frac{\pi}{R_{\text{out}} - R_{\text{in}}} \\
\xi &=& \tan\left(k R_{\text{out}}\right)
\end{eqnarray}
The relation for the radial wave number $k$ is similar to the relation derived for the spherical quantum dots, only the radius $R$ is replaced by the shell thickness $R_{\text{out}}-R_{\text{in}}$. There is a simple physical interpretation for this relation: the electron will form a standing wave by reflecting between the inner and outer spherical hard-walls, and hence the wave number is inversely proportional to the distance between these walls. The modification to the wave number turns out to be the only change compared to the spherical quantum dots: all quantities are the same for the spherical shell after replacing $R$ by $R_{\text{out}}-R_{\text{in}}$. In other words, the confinement energy $\lambda$ and magnetic moment are parameterized by the radial wave number $k$. The magnetic moment depends therefore one-to-one on the confinement energy and it is not possible to tune the magnetic moment and the confinement energy of the state separately. We exemplify this in Fig.~\ref{fig:mu_shell}, where we show the integrated orbital moment ${\boldsymbol \mu}_{\text{IC-BV}}$ of InAs shells for various ratios of $R_{\text{in}}/R_{\text{out}}$ as function of the confinement energy: all curves fall on top of each other. We will show in section \ref{sec:cylindrical} that it is possible to independently tune the magnetic moment and confinement energy if the symmetry of the nanostructure is lowered. Of course it is possible to tune the magnetic moment by changing the shell thickness (see Fig.~\ref{fig:mu_shell}), which can either be done structurally (e.g. in colloidal quantum dots), or electrically by using gates. 

In Fig.~\ref{fig:mu_spatial_shell}(a) we plot the spatial distribution of the dominant current density $\langle {\bf j} \rangle^{\text{BV}}$ for $R_{\text{in}}/R_{\text{out}}=\tfrac{1}{2}$. The current distribution consists of an inner and outer current loop, propagating in opposite directions and partially canceling each other. These currents create oppositely oriented orbital moments, which can also be seen directly from the expression for the dominant integrated orbital moment ${\boldsymbol \mu}_{\text{IC-BV}}$:
\begin{eqnarray}
{\boldsymbol \mu}_{\text{IC-BV}} &=& -\mu_B\sqrt{\frac{2}{3}}\frac{m_0 P_0}{\pi\hbar^2} \frac{\text{Im}\{\alpha - \sqrt{2}\beta\}}{1+|\alpha|^2+|\beta|^2} \left(R_{\text{out}} - R_{\text{in}}\right) {\bf e}_z \nonumber\\
\end{eqnarray}
which can be interpreted as the difference between the orbital moment generated by the outer current and the inner current. The degree of cancelation depends therefore on the shell thickness, which shows again that the orbital moment for spherical systems is uniquely determined by the radial wave number $k$.

The current distribution of spherical shells are markedly different from the current distribution of the spheres. Removal of material from the center of the spheres has therefore a non-trivial effect on the orbital currents: the number of circulating currents and the direction in which they circulate seems to be linked to the topology (genus) of the nanostructure. This resembles the current distributions in the quantum spin hall effect\cite{Kane2005} and this analogy will be subject of further study.

\section{Cylindrical symmetry} \label{sec:cylindrical}

In the previous section we found that the nanostructures with spherical symmetry are governed by a single geometrical parameter. By reducing the symmetry of the confining potential to cylindrical symmetry, we can investigate if shape anisotropy can add a new geometrical handle on the magnetic moment. Generally speaking, the ${\bf k}\cdot{\bf p}$ Hamiltonian ${\cal H}$ can formally be decomposed in terms having respectively cylindrical, cubic and tetragonal symmetry~\cite{Winkler2003}:
\begin{eqnarray}
{\cal H} = {\cal H}_{\text{cyl}} + {\cal H}_{\text{cub}} + {\cal H}_{\text{tet}}
\end{eqnarray}
Since we will be investigating cylindrically symmetric nanostructures, we will use only the cylindrically symmetric part ${\cal H}_{\text{cyl}}$. Moreover, it has been shown\cite{Trebin1979} that ${\cal H}_{\text{cub}}$ is proportional to $\gamma_3-\gamma_2$, which is for most semiconductors a small quantity compared to $\gamma_1$ and $\gamma_2$ (see Table~\ref{table:materials}). Analogous to Sec.~\ref{sec:spherical}, this Hamiltonian will now be block diagonal in a basis of eigenstates of $F_z$\cite{Sercel1990}:
\begin{eqnarray}
{\cal H} = \sum_{F_z} {\cal H}_{F_z}
\end{eqnarray}
since only $F_z=L_{\text{E},z}+J_z$, the projection of the total angular momentum on the symmetry axis, remains quantized for nanostructures with cylindrical symmetry. A convenient basis are the product states:
\begin{eqnarray}
|F_z;J,J_z;k,k_z\rangle = |J,J_z\rangle |k,k_z,L_{\text{E},z}=F_z-J_z\rangle
\end{eqnarray}
where $|J,J_z\rangle$ are Bloch functions, $|k,k_z,L_{\text{E},z}=F_z-J_z\rangle$ the envelope wave functions, $k$ is the radial wave number, and $k_z$ the wave number along the symmetry axis (which we choose to be the $z$-axis). The envelope wave function has the coordinate representation:
\begin{eqnarray}
&& \langle r,\theta,z |k,k_z,L_{\text{E},z}=F_z-J_z\rangle = \\
&& \hspace{10mm} \frac{i^{L_{\text{E},z}}}{2\pi} \left\{J_{L_{\text{E},z}}\left(kr\right)+\xi N_{L_{\text{E},z}}\left(kr\right)\right\} e^{i L_{\text{E},z} \theta} e^{i k_z z}\nonumber
\end{eqnarray}
where $J_l(r)$ is the $l$th-order Bessel function of the first kind, $N_l(r)$ is the $l$th-order Neumann function of the first kind, and $\xi$ a dimensionless parameter determined by the boundary conditions. Using the transformation as outlined in Ref.~\onlinecite{Sercel1990}, we can represent ${\cal H}_{\text{cyl}}$ in the cylindrical envelope basis. The resulting Hamiltonian is shown in Table~\ref{table:Hcyl}; the basis of Bloch functions can be found in Table~\ref{table:Bloch}. We would like to point out that, although the transformation of Ref.~\onlinecite{Sercel1990} is correct, the cylindrical symmetry is not correctly introduced in their Hamiltonian. We have therefore used the correctly derived Hamiltonian of Ref.~\onlinecite{Trebin1979}. The Hamiltonian of Ref.~\onlinecite{Sercel1990} and our Hamiltonian are identical in the spherical approximation ($\gamma_2 = \gamma_3 = \gamma_{23}$, where $\gamma_{23} = \frac{2}{5}\gamma_2 + \frac{3}{5}\gamma_3$). We will show that only in the cylindrical approximation it will be possible to independently tune the confinement energy and magnetic moment.

\subsection{Disks with hard-wall boundaries} \label{sec:harddisks}

The confining potential of a disk with radius $R$ and height $H$ with hard-wall boundaries is given by:
\begin{eqnarray}
V(r,z) = \left\{
\begin{array}{lc}
0 & r \leq R \quad \text{and} \quad |z|\leq H/2 \\
\infty & \text{elsewhere}
\end{array}
\right.
\end{eqnarray}
The envelope wave function needs to be normalizable at the center of the disk, hence only Bessel functions $J_l(kr)$ contribute to the envelope wave function (i.e. $\xi=0$). Furthermore, the traveling wave $e^{ik_z z}$ in the $z$-direction will become a standing wave. Since we assume that the electron ground state predominantly originates from conduction band states, we choose the approximate boundary condition:
\begin{eqnarray}
\langle r,\theta,z |k,k_z,0\rangle_{r=R,z=\pm\tfrac{H}{2}}=0
\end{eqnarray}
from which the relations $k=\tfrac{\rho_{0,1}}{R}$ and $k_z=\tfrac{\pi}{H}$ follow (where $\rho_{l,m}$ denotes the $m$th zero of the $l$th-order Bessel function). The envelope spinor for the electron ground state with $F_z=+\frac{1}{2}$ becomes then:
\begin{eqnarray*}
\langle r,\theta,z|\Psi\rangle = \frac{N}{2\pi} \left(
\begin{array}{r@{}l}
   v_1 & J_0\left(kr\right) \cos(k_z z) \\
 - v_2 & J_1\left(kr\right) \sin(k_z z) \\
 i v_3 & J_1\left(kr\right) \cos(k_z z) e^{-i\theta} \\
 i v_4 & J_0\left(kr\right) \sin(k_z z) \\
 i v_5 & J_1\left(kr\right) \cos(k_z z) e^{+i\theta} \\
-i v_6 & J_2\left(kr\right) \sin(k_z z) e^{+2i\theta} \\
 i v_7 & J_0\left(kr\right) \sin(k_z z) \\
 i v_8 & J_1\left(kr\right) \cos(k_z z) e^{+i\theta}
\end{array}
\right)
\hspace{2mm}
\begin{array}{c}
\text{CB$\uparrow$} \\
\text{CB$\downarrow$} \\
\text{HH$\uparrow$} \\
\text{LH$\uparrow$} \\
\text{LH$\downarrow$} \\
\text{HH$\downarrow$} \\
\text{SO$\uparrow$} \\
\text{SO$\downarrow$} 
\end{array} \label{eq:psi_cyl}
\end{eqnarray*}
where the coefficients $v_i$ indicate the amount of intermixing of different Bloch states (comparable to $\alpha$ and $\beta$ of the spheres), and where $N$ is a normalization constant:
\begin{eqnarray}
|N|^2 = \frac{8\pi\rho_{0,1}^2}{H R^2J_1(\rho_{0,1})^2 \left(\sum_{i=1}^8 |v_i|^2 \rho_{0,1}^2 - 4 |v_6|^2\right)}
\end{eqnarray}
We will first investigate the composition of the ground state, which in general depends both on the radius and height of the disk. As the composition depends on the coefficients $v_i$, we need to diagonalizing ${\cal H}_{\text{cyl}}$ to find their analytical expressions. Unfortunately these expressions are rather cumbersome, and it proves more insightful to analyze the coefficients in the quantum well limit (QW), for which $k=0$:
\begin{widetext}
\begin{eqnarray}
v_1^{\text{QW}} = 1 \quad\quad v_2^{\text{QW}} = 0 \quad\quad v_3^{\text{QW}} = 0 \quad\quad v_5^{\text{QW}} = 0 \quad\quad v_6^{\text{QW}} = 0 \quad\quad v_8^{\text{QW}} = 0
\end{eqnarray}
\begin{eqnarray}
v_4^{\text{QW}} &=& \frac{i\sqrt{\frac{2}{3}}k_z P_0\left[\frac{\hbar^2k_z^2}{2m_0}(\gamma_1-2\gamma_2)+(E_g+\Delta+\lambda)\right]}{\left[\frac{\hbar^2k_z^2}{2m_0}(\gamma_1-2\gamma_2)+(E_g+\Delta+\lambda)\right]\left[\frac{\hbar^2k_z^2}{2m_0}(\gamma_1+4\gamma_2)+(E_g+\lambda)\right] - 2\frac{\hbar^2k_z^2}{2m_0}\gamma_2 \Delta} \\
v_7^{\text{QW}} &=& \frac{i\sqrt{\frac{1}{3}}k_z P_0\left[\frac{\hbar^2k_z^2}{2m_0}(\gamma_1-2\gamma_2)+(E_g+\lambda)\right]}{\left[\frac{\hbar^2k_z^2}{2m_0}(\gamma_1-2\gamma_2)+(E_g+\Delta+\lambda)\right]\left[\frac{\hbar^2k_z^2}{2m_0}(\gamma_1+4\gamma_2)+(E_g+\lambda)\right] - 2\frac{\hbar^2k_z^2}{2m_0}\gamma_2 \Delta}
\end{eqnarray}
and in the nanowire limit (NW), for which $k_z=0$:
\begin{eqnarray}
v_1^{\text{NW}} = 1 \quad\quad v_2^{\text{NW}} = 0 \quad\quad v_4^{\text{NW}} = 0 \quad\quad v_6^{\text{NW}} = 0 \quad\quad v_7^{\text{NW}} = 0
\end{eqnarray}
\begin{eqnarray}
v_3^{\text{NW}} &=& \frac{i\sqrt{2}k P_0\left[K_1+(E_g+\Delta+\lambda)\right]\left[K_2+2(E_g+\lambda)\right]}{\left[K_1+(E_g+\Delta+\lambda)\right]\left[K_2+2(E_g+\lambda)\right]\left[K_3+2(E_g+\lambda)\right] - 8\frac{\hbar^2k^2}{2m_0}\gamma_2 \Delta\left[\frac{\hbar^2k^2}{2m_0}(\gamma_1+\gamma_2)-\frac{\hbar^2k^2}{2m_0}\frac{3(\gamma_2+\gamma_3)^2}{4\gamma_2}+(E_g+\lambda)\right]} \nonumber\\ \\
v_5^{\text{NW}} &=& \frac{i\sqrt{\frac{2}{3}}k P_0\left[K_1+(E_g+\Delta+\lambda)\right]\left[K_2+2(E_g+\lambda)\right]}{\left[K_1+(E_g+\Delta+\lambda)\right]\left[K_2+2(E_g+\lambda)\right]\left[K_3+2(E_g+\lambda)\right] - 8\frac{\hbar^2k^2}{2m_0}\gamma_2 \Delta\left[\frac{\hbar^2k^2}{2m_0}(\gamma_1+\gamma_2)-\frac{\hbar^2k^2}{2m_0}\frac{3(\gamma_2+\gamma_3)^2}{4\gamma_2}+(E_g+\lambda)\right]} \nonumber\\ \\
v_8^{\text{NW}} &=& \frac{i\sqrt{\frac{4}{3}}k P_0\left[K_1+(E_g+\lambda)\right]\left[K_2+2(E_g+\lambda)\right]}{\left[K_1+(E_g+\Delta+\lambda)\right]\left[K_2+2(E_g+\lambda)\right]\left[K_3+2(E_g+\lambda)\right] - 8\frac{\hbar^2k^2}{2m_0}\gamma_2 \Delta\left[\frac{\hbar^2k^2}{2m_0}(\gamma_1+\gamma_2)-\frac{\hbar^2k^2}{2m_0}\frac{3(\gamma_2+\gamma_3)^2}{4\gamma_2}+(E_g+\lambda)\right]}\nonumber\\ 
\end{eqnarray}
\end{widetext}
where $\lambda$ is the confinement energy following from one of the roots of $|{\cal H}_{\text{cyl}}-\lambda I|=0$, and where
\begin{eqnarray}
K_1 &=& \frac{\hbar^2k^2}{2m_0}(\gamma_1-2\gamma_2) \\
K_2 &=& \frac{\hbar^2k^2}{2m_0}(2\gamma_1-\gamma_2-3\gamma_3) \\
K_3 &=& \frac{\hbar^2k^2}{2m_0}(2\gamma_1+5\gamma_2+3\gamma_3)
\end{eqnarray}
It is clear that only very specific valence bands are mixing into the electron ground state for quantum wells and nanowires, which can be explained as follows. The ground state has only a finite envelope momentum associated with the directions in which the state is confined (at zero temperature). For example, in a quantum well there is only an envelope momentum in the $z$-direction ($k_z\neq0$), since there will be no motion in the plane ($k=0$). In a ${\bf k}\cdot{\bf p}$-model, the envelope momentum ${\bf k}$ is coupled to the atomic orbitals of the crystal ${\bf p}$. This means that only valence band Bloch states with atomic orbitals which are oriented in the confined directions will participate in the ground state. Thus for the quantum well, only valence band Bloch states with atomic orbitals $|z\rangle$ will contribute, while for the nanowire only the atomic orbital states $|x\rangle$ or $|y\rangle$ are relevant (see Table~\ref{table:Bloch}). Since we are examining the $F_z=+\tfrac{1}{2}$ ground state, only the spin $\uparrow$-part of the valence band Bloch state can participate. Hence only $\{$LH$\uparrow,$ SO$\uparrow\}$ (or $v_{4,7}^{\text{QW}}$) mix into the ground state of the quantum well, while only  $\{$HH$\uparrow,$ LH$\downarrow,$ SO$\downarrow\}$ (or $v_{3,5,8}^{\text{NW}}$) are relevant for the nanowire.

\begin{figure}
\begin{center}
\includegraphics[width=\columnwidth]{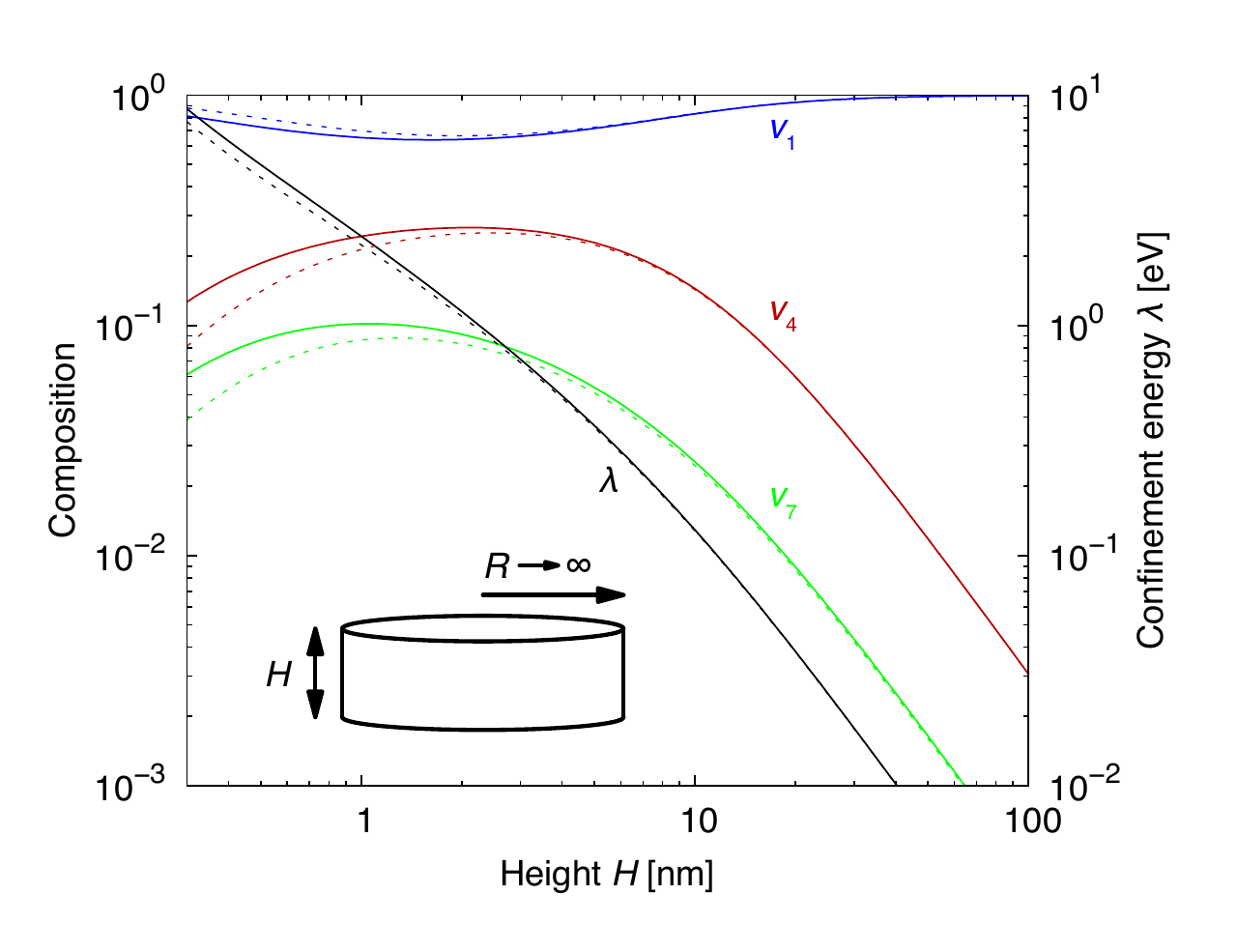}
\caption{The height dependence of the composition and confinement energy $\lambda$ of an InAs quantum well, where only $v_{4,7}$ intermix, in the cylindrical approximation (continuous lines) and in the spherical approximation (dotted lines).}
\label{fig:mix_QW}
\end{center}
\end{figure}

\begin{figure}
\begin{center}
\includegraphics[width=\columnwidth]{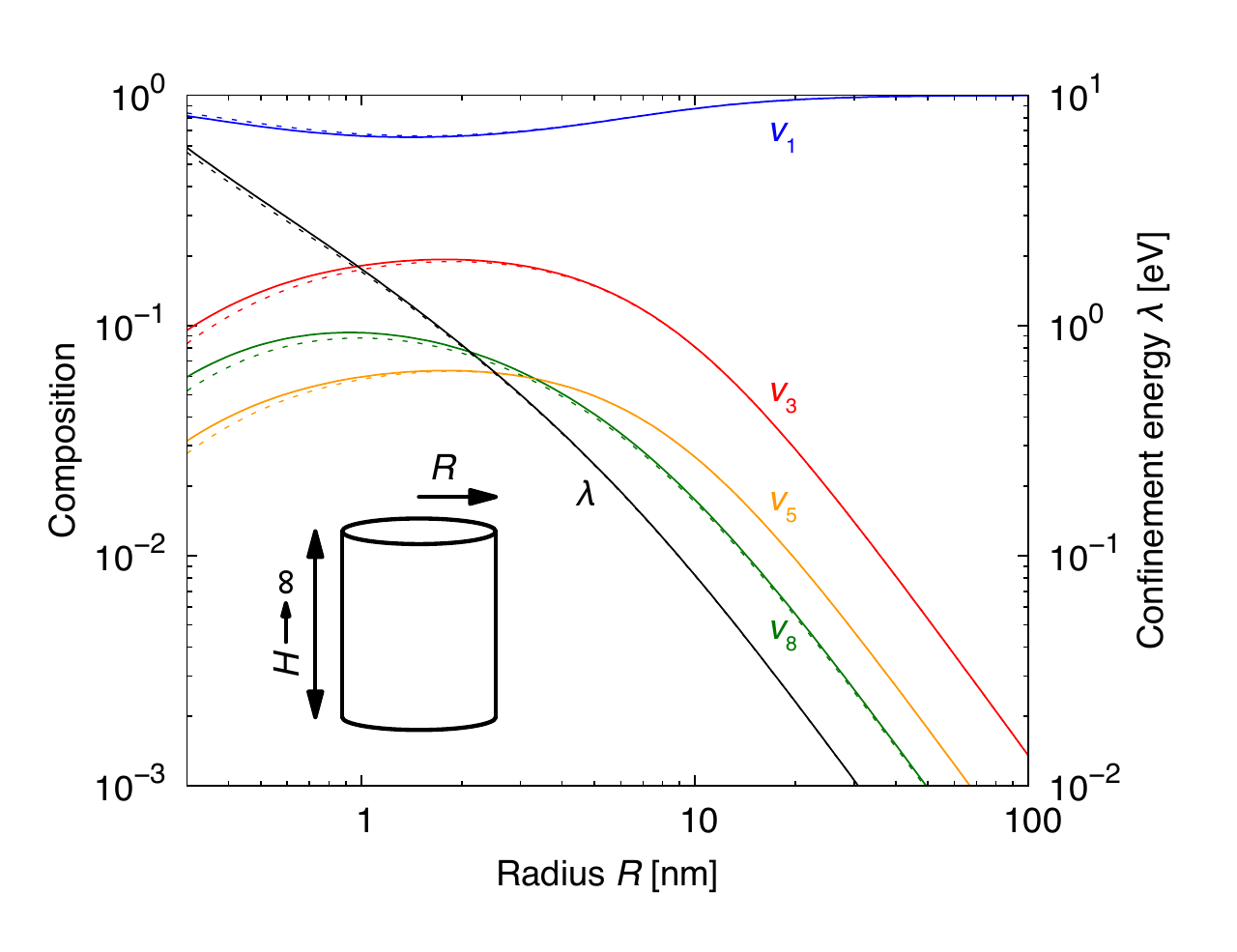}
\caption{The radius dependence of the composition and confinement energy $\lambda$ of an InAs nanowire, where only $v_{3,5,8}$ intermix, in the cylindrical approximation (continuous lines) and in the spherical approximation (dotted lines).}
\label{fig:mix_NW}
\end{center}
\end{figure}

In Fig.~\ref{fig:mix_QW}~and~\ref{fig:mix_NW} we show the height (radius) dependence of the composition of an InAs quantum well and nanowire. The composition behaves qualitatively the same as for the spheres: the intermixing of valence band states peaks at a certain height (or radius) and the conduction band contribution has a minimum of $\sim65\%$. The reason for this behavior is also the same and can be directly observed in the expressions for the $v_i$: there is a competition between the coupling term ($k P_0$ or $k_z P_0$) and the free kinetic energy ($\propto k^2$ or $k_z^2$). We would like to point out that the exact dependence of the coefficients on the height (or radius) is determined by the free kinetic energies associated with the confined direction. In particular, coefficients $v_{4,7}^{\text{QW}}$ depend only on combinations of $\gamma_1$ and $\gamma_2$ which represent the effective hole masses along the $z$-direction. However, the coefficients $v_{3,5,8}^{\text{NW}}$ for a nanowire also depend on $\gamma_3$, as the combinations of $\gamma$'s involve the in-plane effective hole masses. When using the spherical approximation these differences disappear and all coefficients have the same functional dependence on radius or height, as can be seen by comparing the dotted lines of Fig.~\ref{fig:mix_QW}~and~\ref{fig:mix_NW}.

\begin{figure*}
\begin{center}
\includegraphics[width=2\columnwidth]{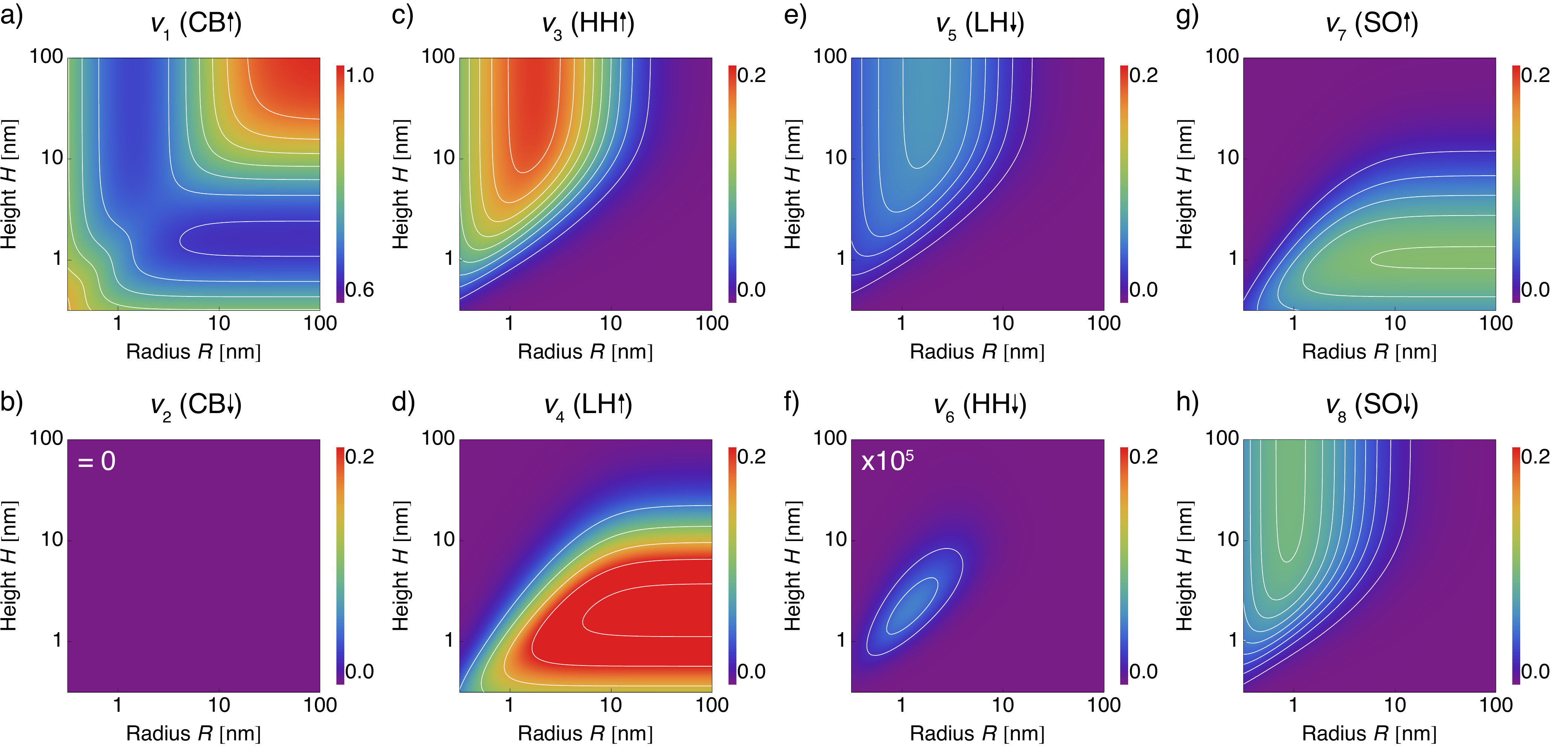}
\caption{(a-h) The height and radius dependence of the composition of an InAs disk. The white lines are contours constant $|N|^2|v_i|^2$. The electron ground state $F_z=+\tfrac{1}{2}$ is dominated by the $v_1$ (CB$\uparrow$) state; note that (a) has a different color scale than the other plots. In the quantum wire limit, i.e. very large $H$, only $v_3$ (HH$\uparrow$), $v_5$ (LH$\downarrow$) and $v_8$ (SO$\downarrow$) mix into the ground state. Conversely, in the quantum well limit, i.e. very large $R$, only $v_4$ (LH$\uparrow$) and $v_7$ (SO$\uparrow$) mix into the ground state. The contributions of $v_2$ (CB$\downarrow$) is absent and of $v_6$ (HH$\downarrow$) negligible (enhanced by $10^5$ for visibility).}
\label{fig:mix_cylinder}
\end{center}
\end{figure*}

In Fig.~\ref{fig:mix_cylinder} we show the radius and height dependence of the composition of a finite InAs disk. As expected, the electron ground state $F_z=+\tfrac{1}{2}$ is always dominated ($\geq65\%$) by the CB$\uparrow$ states. The radius and height dependence of coefficients $v_{4,7}$ is similar to coefficients $v_{3,5,8}$, only the roles of radius and height are interchanged. We therefore discuss only the dependence of coefficients $v_{3,5,8}$. Of course only coefficients $v_{3,5,8}$ play a role if the height is very large, since we are then approaching the nanowire limit. Moreover, the non-monotonic dependence of these coefficients on the radius follows the explanation of the previous paragraph. To understand why they have only a significant weight in a triangular region of the $RH$-space, we need to analyze their less-intuitive dependence on the height. To first order, we can use the analytical expressions for $v_{3,5,8}^{\text{NW}}$, recognizing that the confinement energy $\lambda$ depends in general on both the radius and height. Indeed, in the limit of large radius and height, $\lambda$ becomes:
\begin{eqnarray}
\lim_{R,H\rightarrow\infty} \lambda = \frac{E_g + \tfrac{2}{3}\Delta}{E_g(E_g + \Delta)}(k^2 + k_z^2) P_0^2
\end{eqnarray}
Inserting this expression for $\lambda$ into $v_{3,5,8}^{\text{NW}}$, we see that $v_{3,5,8} \sim 1/k_z^2$, and will thus decrease monotonically when the height gets smaller. This effect is only significant when the confinement energy $(\propto k_z^2)$ is comparable to the free kinetic energies $(\propto k^2)$, i.e. when the $H\sim R$. When $H\ll R$, the confinement energy has quenched the coefficients $v_{3,5,8}$ completely, which explains the insignificance of these coefficients in the triangular region of the $RH$-space.

We find that $v_2$ is always zero, even though there are second-order couplings between CB$\downarrow$ and CB$\uparrow$ (see Table~\ref{table:Hcyl}). It turns out that these couplings are canceling each other in the cylindrical approximation. We expect that this is no longer true when cubic terms are included in the Hamiltonian. Coefficient $v_6$, however, still has a finite weight due to third-order couplings, since the HH$\downarrow$ Bloch state can only couple to the CB$\uparrow$ Bloch state via two intermediate valence band Bloch states. This explains why $v_6$ has a (extremely) small weight in a very limited region of the $RH$-space, as it depends on the overlap of coefficients $v_{3,5,8}$ and $v_{4,7}$. Interestingly, we find that $v_6 \propto (\gamma_2 - \gamma_3)$ and consequently $v_6=0$ in the spherical approximation. Due to the extremely small weight, we set $v_6=0$ to simplify the analytical expressions.

\begin{figure*}
\begin{center}
\includegraphics[width=2\columnwidth]{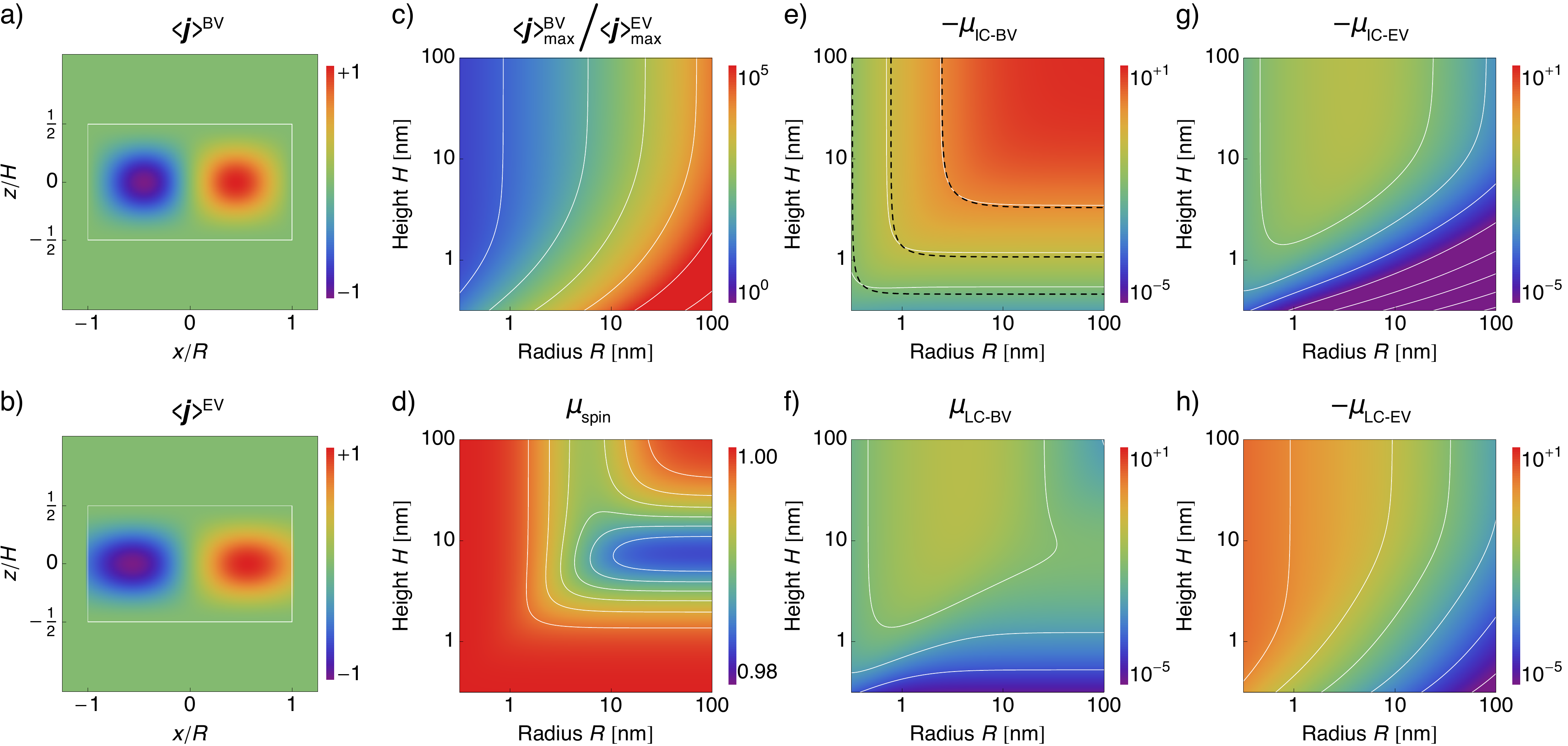}
\caption{(a-b) The spatial distribution of the normalized magnitude of the ${\bf e}_y$-component of $\langle {\bf j} \rangle^{\text{BV}}$ (a) and $\langle {\bf j} \rangle^{\text{EV}}$ (b) of a disk, with $v_{6}=0$. The current density peaks at roughly $R/2$ and $z=0$ and resembles a current loop in the plane of the disk. (c) The height and radius dependence of the ratio between the peak current densities $\langle {\bf j} \rangle^{\text{BV}}_{\text{max}}$ and $\langle {\bf j} \rangle^{\text{EV}}_{\text{max}}$ for an InAs disk. The color scale is logarithmic and the leftmost contour indicates a ratio of 10. As long as $R\geq1$~nm, $\langle {\bf j} \rangle^{\text{BV}} \gg \langle {\bf j} \rangle^{\text{EV}}$. (d-h) The height and radius dependence of the different integrated orbital momenta (e-h) and the integrated spin moment (d) for an InAs disk, all in units of Bohr magnetons. The color scale is logarithmic and the white lines are contours of constant moment with a power of 10. The dashed black lines in (e) are contours of constant confinement energy.}
\label{fig:j_mu_cylinder}
\end{center}
\end{figure*}

The itinerant BV related current density $\langle {\bf j} \rangle^{\text{BV}}$ of the electron ground state can be expressed in terms of coefficients $v_i$:
\begin{eqnarray}
&&\hspace{-4mm}\langle {\bf j} \rangle^{\text{BV}} = -\frac{e |N|^2 P_0}{2 \sqrt{6} \pi ^2 \hbar } \big[ \\
\hspace{2mm}& & J_0(kr) J_1(kr) \cos^2(k_z z) v_1 \text{Im}\{\sqrt{3}v_3 + v_5 - \sqrt{2}v_8\} \big] {\bf e}_{\theta}\nonumber
\end{eqnarray}
The current is flowing in the ${\bf e}_{\theta}$-direction and the radial distribution is governed by the product of the conduction and valence band envelope wave functions, i.e. $J_0(kr)J_1(kr)$. This resembles again a current loop, see Fig.~\ref{fig:j_mu_cylinder}(a), peaking at about $R/2$ and $z=0$. The momentum matrix elements associated with the ${\bf e}_{\theta}$-direction can be written as $\langle u_j | \tfrac{1}{\rho} \tfrac{i}{\hbar} L_{\text{B},z} | u_i \rangle$. The current depends indeed on coefficients $v_{3,5,8}$, which represent valence band Bloch states carrying an orbital momentum $L_{\text{B},z}=\pm1$. In the limit of large radius and height, we find that the particular combination of these coefficients is proportional to the spin-orbit coupling:
\begin{eqnarray}
&&\hspace{-10mm}\lim_{R,H\rightarrow\infty} \text{Im}\{\sqrt{3}v_3 + v_5 - \sqrt{2}v_8\} \\
&\hspace{4mm}=& \sqrt{\frac{3}{2}} \frac{k P_0}{E_g + \lambda} - \sqrt{\frac{1}{6}} \frac{k P_0}{E_g + \lambda} - \sqrt{\frac{2}{3}} \frac{k P_0}{E_g + \Delta + \lambda}\nonumber\\ \\
&\hspace{4mm}=& \sqrt{\frac{2}{3}} k P_0 \frac{\Delta}{(E_g + \lambda)(E_g+\Delta+\lambda)}
\end{eqnarray}
which shows explicitly the spin-orbit correlated nature of the current. It also shows explicitly the cancellation mechanism, as discussed for the spheres: coefficient $v_3$ and coefficients $v_{5,8}$ create oppositely circulating currents, since they have respectively $L_{\text{B},z}=+1$ and $L_{\text{B},z}=-1$. The degree of cancellation depends on the spin-orbit splitting $\Delta$, as this tunes the presence of the SO$\downarrow$ ($v_8$) Bloch state. The proportionality to $k$ suggests that the current would be quenched in the quantum well limit. We will show later on that this does not mean that the orbital moment ${\boldsymbol \mu}_{\text{IC-BV}}$ associated with this current is quenched in quantum wells. Furthermore we point out that quenching of $\langle {\bf j} \rangle^{\text{BV}}$ in the quantum well limit can only happen in a perfect crystal at a temperature of zero Kelvin. In practice, either the finite temperature (the de Broglie wavelength) or dopants (the Bohr radius) will lead to a finite radial wave number and therefore to a finite current.

The itinerant EV related current density $\langle {\bf j}\rangle^{\text{EV}}$ can be more generally calculated using the general envelope state $|k,k_z,L_{\text{E},z}\rangle$:
\begin{eqnarray}
\langle {\bf j} \rangle^{\text{EV}}_{|k,k_z,L_{\text{E},z}\rangle} = \frac{e \hbar}{4 \pi^2 r m_0} L_{\text{E},z} J_{L_{\text{E},z}}(kr)^2 {\bf e}_{\theta}
\end{eqnarray}
where we have left out the ${\bf e}_z$-component, since for the disk the traveling plane wave $e^{i k_z z}$ is replaced by a standing wave, which cannot carry a current. Once more, it is clear the $\langle {\bf j} \rangle^{\text{EV}}$ flows in the ${\bf e}_{\theta}$-direction and is generated by envelope wave functions having a finite $L_{\text{E},z}$, meaning that the dominant CB$\uparrow$ Bloch state will not contribute. The itinerant EV related current density $\langle {\bf j} \rangle^{\text{EV}}$ of the ground state can also be expressed in terms of coefficients $v_i$:
\begin{eqnarray}
&&\hspace{-10mm} \langle {\bf j} \rangle^{\text{EV}} = -\frac{e|N|^2 \hbar}{4 \pi^2 r m_0} \big[ \\
\hspace{6mm}&& J_1(kr)^2 \cos^2(k_z z) \left\{|v_3|^2-|v_5|^2-|v_8|^2\right\} \big] {\bf e}_{\theta}\nonumber
\end{eqnarray}
It is clear that this current has the same spatial symmetry as $\langle {\bf j} \rangle^{\text{BV}}$ and resembles a current loop. The radial distribution is slightly different, being proportional to $J_1(kr)^2$, as can also be seen in Fig.~\ref{fig:j_mu_cylinder}(b). It is straightforward to show that $\langle {\bf j} \rangle^{\text{EV}}$ is also proportional to the spin-orbit coupling. The Bloch velocity related current dominates the envelope velocity related current, as can be seen from Fig.~\ref{fig:j_mu_cylinder}(c). For realistic sizes (i.e. $R\geq 1$~nm), the peak current density $\langle {\bf j} \rangle^{\text{BV}}_{\text{max}}$ is more than 10 times larger than $\langle {\bf j} \rangle^{\text{EV}}_{\text{max}}$.

Now that the currents are know, we can analyze the magnetic moments. Since the spatial symmetries of the orbital moments and spin moment are similar to the case of the spheres, we do not discuss them in detail. Instead we will focus on the integrated moments; the integrated orbital moments are:
\begin{eqnarray}
{{\boldsymbol \mu}_{\text{IC-BV}}} &=& -{\mu_B}\sqrt{\frac{2}{3}} \frac{m_0 P_0 R}{\hbar^2 \rho_{0,1}} \frac{v_1 \text{Im}\{\sqrt{3} v_3 + v_5 - \sqrt{2} v_8\}}{\sum_{i=1}^8 |v_i|^2} {\bf e}_z \label{eq:muICBV_disk} \\
{{\boldsymbol \mu}_{\text{LC-BV}}} &=& {\mu_B}\frac{|v_3|^2 - \frac{1}{3} \text{Im}\{v_5 - \sqrt{2} v_8\}^2 + \frac{1}{3} \text{Im}\{v_4 + \sqrt{2} v_7\}^2}{\sum_{i=1}^8 |v_i|^2} {\bf e}_z  \nonumber \\ \label{eq:muLCBV_disk} \\
{{\boldsymbol \mu}_{\text{IC-EV}}} &=& -{\mu_B}\frac{|v_3|^2 - |v_5|^2 - |v_8|^2}{\sum_{i=1}^8 |v_i|^2} {\bf e}_z \label{eq:muICEV_disk} \\
{{\boldsymbol \mu}_{\text{LC-EV}}} &=& -{\mu_B}\sqrt{\frac{2}{3}}\frac{P_0 \rho_{0,1}}{R} \frac{v_1 \text{Im} \{\frac{\sqrt{3}}{E_g} v_3 + \frac{1}{E_g} v_5 - \frac{\sqrt{2}}{E_g+\Delta} v_8\}}{\sum_{i=1}^8 |v_i|^2} {\bf e}_z \nonumber \\
\end{eqnarray}
while the integrated spin moment becomes:
\begin{eqnarray}
{{\boldsymbol \mu}_{\text{spin}}} &=& {\mu_B}\left[1 - 2 \frac{\frac{1}{3}\text{Im}\{v_4 + \sqrt{2} v_7\}^2 + \frac{1}{3}\text{Im}\{\sqrt{2} v_5 + v_8\}^2}{\sum_{i=1}^8 |v_i|^2}\right] {\bf e}_z \nonumber \\
\end{eqnarray}
As expected for the $F_z=+\tfrac{1}{2}$ ground state, the integrated magnetic moments are oriented along the ${\bf e}_z$-direction. In Fig.~\ref{fig:j_mu_cylinder}(d-h) we plot the radius and height dependence of these magnetic moments for an InAs disk.

\begin{figure}
\begin{center}
\includegraphics[width=\columnwidth]{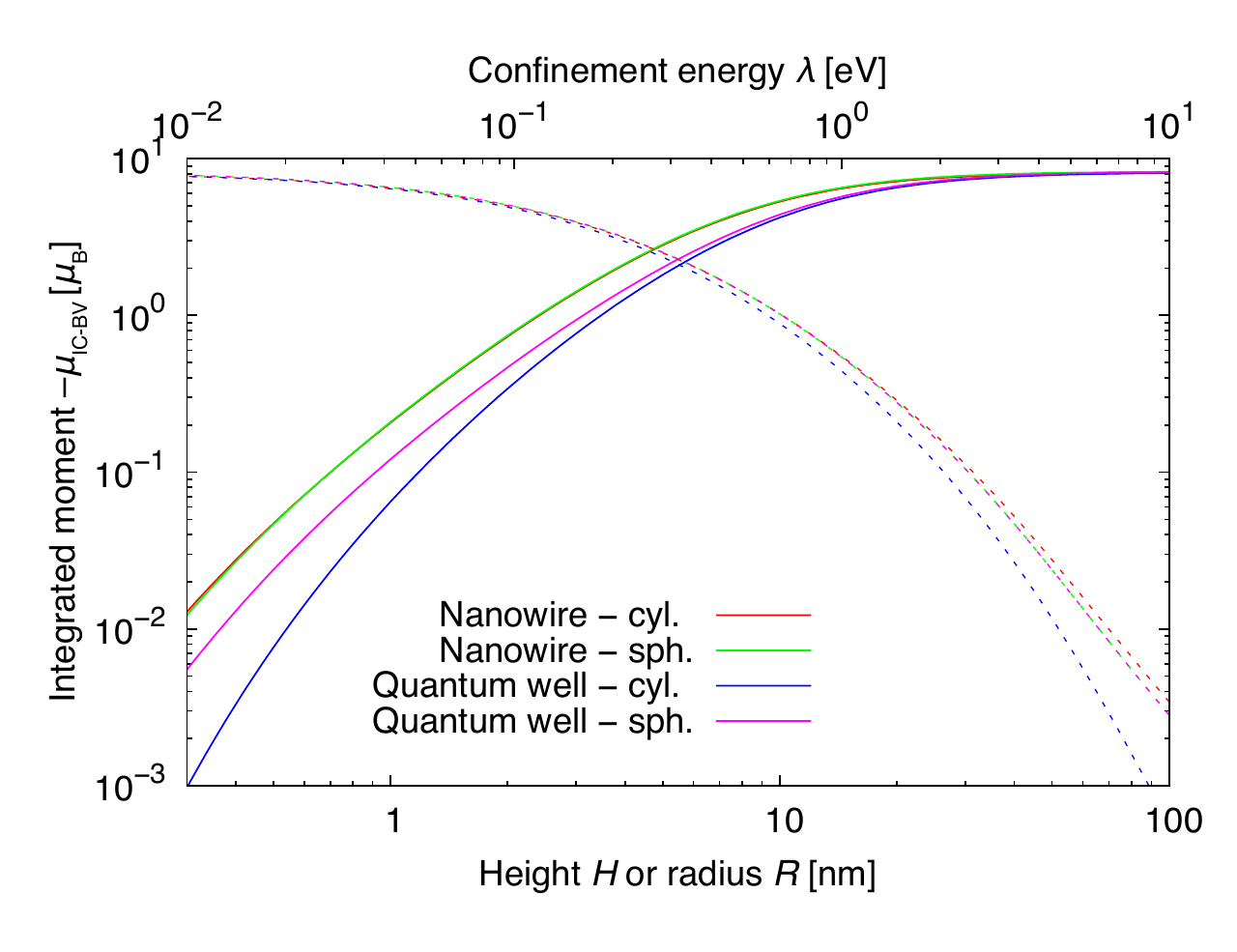}
\caption{The integrated orbital moment ${\boldsymbol \mu}_{\text{IC-BV}}$ of an InAs quantum well/nanowire as function of height/radius (continuous lines) and confinement energy (dotted lines).}
\label{fig:mu_QNW}
\end{center}
\end{figure}

Similar to the case of the spheres, ${\boldsymbol \mu}_{\text{IC-BV}}$ dominates over all other orbital moments within the range of the validity of the envelope function approximation. It is interesting to examine the behavior of ${\boldsymbol \mu}_{\text{IC-BV}}$ in several limiting cases. In the limit of large radius and height ${\boldsymbol \mu}_{\text{IC-BV}}$ becomes again the Roth formula\cite{Roth1959}, and the disk behaves as a bulk material. In the limit of large height, we can examine the behavior of ${\boldsymbol \mu}_{\text{IC-BV}}$ in nanowires. In Fig.~\ref{fig:mu_QNW} we show the radius dependence of ${\boldsymbol \mu}_{\text{IC-BV}}$ of an InAs nanowire. The origin of this dependence is the same as we have found for the spheres, and can be explained using the simple current loop: the non-monotonic radius dependence of the integrated current is multiplied by $R^2$. This makes the orbital moment constant at large radius, and depend on $R^4$ for small radius. In the limit of large radius, we can examine the behavior of ${\boldsymbol \mu}_{\text{IC-BV}}$ in quantum wells. Since ${\boldsymbol \mu}_{\text{IC-BV}}$ originates from $\langle {\bf j} \rangle^{\text{BV}}$, this moment is proportional to coefficients $v_{3,5,8}$. As was pointed out earlier, these coefficients are proportional to $k$ in the limit of large radius, so one might expect ${\boldsymbol \mu}_{\text{IC-BV}}$ to quench in the quantum well limit. However, the orbital moment is proportional to ${\bf r} \times {\bf j}$. Since ${\bf r}\sim 1/k$, the lever arm ${\bf r}$ cancels the $k$-dependence of the current ${\bf j}$. Therefore ${\boldsymbol \mu}_{\text{IC-BV}}$ remains non-zero in the quantum well limit. We show the height dependence of ${\boldsymbol \mu}_{\text{IC-BV}}$ for an InAs quantum well in Fig.~\ref{fig:mu_QNW}. The quenching mechanism is slightly different from that of the nanowires: the height enters ${\boldsymbol \mu}_{\text{IC-BV}}$ mainly through the confinement energy. To show these differences more clearly, we can investigate the analytical expressions for ${\boldsymbol \mu}_{\text{IC-BV}}$ in nanowires (quantum wells), in the limit of large radius (height):
\begin{widetext}
\begin{eqnarray}
\frac{{\boldsymbol \mu}_{\text{IC-BV}}^{\text{NW}}}{\mu_{\text{Roth}}} &=& 1 - \left(\frac{2(\gamma_1+\gamma_2)}{E_g} + \frac{2\frac{m_0P_0^2}{\hbar^2} - \gamma_1 \Delta}{E_g(E_g+\Delta)} + 2\frac{m_0P_0^2}{\hbar^2}\frac{6(E_g+\frac{2}{3}\Delta)^2}{E_g^2(E_g+\Delta)^2}\right)\frac{\hbar^2 k^2}{2m_0} + {\cal O}(k^3) \\
\frac{{\boldsymbol \mu}_{\text{IC-BV}}^{\text{QW}}}{\mu_{\text{Roth}}} &=& 1 - \left(\frac{2(\gamma_1-2\gamma_2+3\gamma_3)}{E_g} + \frac{2\frac{m_0P_0^2}{\hbar^2} - \gamma_1 \Delta}{E_g(E_g+\Delta)} + 2\frac{m_0P_0^2}{\hbar^2}\frac{6(E_g+\frac{2}{3}\Delta)^2}{E_g^2(E_g+\Delta)^2}\right)\frac{\hbar^2 k_z^2}{2m_0} + {\cal O}(k_z^3)
\end{eqnarray}
\end{widetext}
Notice that the functional dependence of ${\boldsymbol \mu}_{\text{IC-BV}}$ is the same for the nanowire and quantum well, except for the particular combination of $\gamma$'s that appears. More precisely, only the quantum well contains contributions from $\gamma_3$, making it prone to changes when going from the cylindrical to the spherical approximation. This is indeed observed for the InAs quantum well in Fig.~\ref{fig:mu_QNW}. In the spherical approximation, we find the orbital moment of nanowires and quantum wells to depend similarly on $k$ and $k_z$. This becomes clear in Fig.~\ref{fig:mu_QNW} when the orbital moment is plotted against the confinement energy: in the spherical approximation the curves of the quantum well and nanowire fall on top of each other. This shows that the orbital moment and confinement energy cannot be tuned independently. In fact, we have checked that both are parameterized by the quantity $k^2+k_z^2$. This behavior is analogous to what has been found for the spherically symmetric nanostructures.

However, when using the cylindrical approximation, the orbital moments of the quantum well and nanowire do not depend in the same way on the confinement energy: in Fig.~\ref{fig:mu_QNW} we see that the confinement energy of a quantum well and nanowire can be the same, yet the orbital moments are different. Indeed we also find that for finite disks it is possible to tune independently the confinement energy and orbital moment, if we use the cylindrical approximation. This is exemplified in Fig.~\ref{fig:j_mu_cylinder}(e), where besides the white lines indicating contours of constant orbital moment, dashed black lines indicate contours of constant confinement energies. It can readily be seen that the two sets of contour lines do not fully overlap, meaning that the confinement energy is changing along a contour line of constant orbital moment. This is distinctively different from nanostructures having spherical symmetry. A cylindrically symmetric nanostructure allows therefore for more versatility in engineering the orbital moment. For example, it is possible to engineer disks with the same confinement energy, yet different orbital momenta. This different behavior arises from the different symmetry of band structure of the crystal and not from the shape of the nanostructure, as we have observed that disks in the spherical approximation are parameterized by $k^2+k_z^2$. The intermixing depends through the confinement energy and the free kinetic energies on the dispersion relation and therefore on the symmetry of the band structure. 

\begin{table}
\caption{The different combinations of the Bloch orbital moment $L_{\text{B},z}$ and envelope orbital moment $L_{\text{E},z}$ for a state having $F_z=+\tfrac{1}{2}$.}
\begin{tabular}{c|r|c|r}
\hline \hline
$v_i$ & $J_z$ & $L_{\text{B},z},s_z$ & $L_{\text{E},z}$ \\
\hline
$v_3$ & $+\tfrac{3}{2}$ & $+1,+\tfrac{1}{2}$ & $-1$ \\
$v_{4,7}$ & $+\tfrac{1}{2}$ & $(+1,-\tfrac{1}{2}), (0,+\tfrac{1}{2})$ & $0$ \\
$v_{5,8}$ & $-\tfrac{1}{2}$ & $(-1,+\tfrac{1}{2}), (0,-\tfrac{1}{2})$ & $+1$ \\
$v_6$ & $-\tfrac{3}{2}$ & $-1,-\tfrac{1}{2}$ & $+2$ \\
\hline \hline
\end{tabular}
\label{table:LB_LE}
\end{table}

The localized orbital moment ${\boldsymbol \mu}_{\text{LC-BV}}$ is proportional to the Bloch orbital moment $\langle u_i | {\bf L}_{\text{B}} | u_j \rangle$. One can recognize in Eq.~\ref{eq:muLCBV_disk} the projection of the orbital Bloch moment $+1$ in front of coefficient $v_3$ (HH$\uparrow$), and $\pm\tfrac{1}{3}$ in front of coefficients $v_{5,8}$ (LH$\uparrow$, SO$\downarrow$) and coefficients $v_{4,7}$ (LH$\downarrow$, SO$\uparrow$). In contrast, ${\boldsymbol \mu}_{\text{IC-EV}}$ is proportional to the envelope orbital moment, since this current originates from $\langle {\bf j} \rangle^{\text{EV}}$. The numerical factors in front of the coefficients in Eq.~\ref{eq:muICEV_disk} are now given by $L_{\text{E},z}$ of the corresponding coefficient. Coefficients $v_{4,7}$ play therefore no role, since the corresponding envelope wave functions have $L_{\text{E},z}=0$. The projection of the Bloch and envelope orbital momenta are related via:
\begin{eqnarray}
F_z &=& L_{\text{E},z} + L_{\text{B},z} + s_z
\end{eqnarray}
In Table~\ref{table:LB_LE} we have tabulated the different possible combinations of the Bloch and envelope orbital momenta for the electron ground state of a disk having $F_z=+\tfrac{1}{2}$. We find that $L_{\text{E},z} = -L_{\text{B},z}$ for coefficients $v_{3,5,8}$. These coefficients dominate the valence band contribution to the electron ground state for $H>R$ (see Fig.~\ref{fig:mix_cylinder}), so that ${\boldsymbol \mu}_{\text{LC-BV}}\approx -{\boldsymbol \mu}_{\text{IC-EV}}$ for $H>R$. The same (near) cancellation in the total orbital moment was found in the spheres. For $H<R$, however, the coefficients $v_{4,7}$ are non-zero so that the cancellation is not so complete. This is different from the spheres and shows again how radius and height have a different influence on the orbital moments in disks.

The localized orbital moment ${\boldsymbol \mu}_{\text{LC-EV}}$ depends via the dipole matrix elements on the momentum matrix elements, see Eq.~\ref{eq:LCEV}. It therefore depends on the same coefficients as ${\boldsymbol \mu}_{\text{IC-BV}}$, although reduced by the band edge energies of the corresponding Bloch states. The main difference between these two moments arises from the different pre-factors. For ${\boldsymbol \mu}_{\text{IC-BV}}$ the pre-factor reflects the lever arm, which is proportional to $R$, while for ${\boldsymbol \mu}_{\text{LC-EV}}$ it reflects the gradient of the envelope wave function, which is proportional to $1/R$ when ${\boldsymbol \mu}_{\text{LC-EV}}$ is oriented along ${\bf e}_z$. This explains why ${\boldsymbol \mu}_{\text{LC-EV}}$ becomes larger when the radius is decreased: the envelope velocity will steadily increase. Note that ${\boldsymbol \mu}_{\text{LC-EV}}$ becomes constant at small radius: coefficients $v_{3,5,8}$ are proportional to $1/k$ due to the free kinetic energy, while the pre-factor depends on $k$, which together makes ${\boldsymbol \mu}_{\text{LC-EV}}$ constant at small radius. By decreasing the height, ${\boldsymbol \mu}_{\text{LC-EV}}$ is quenched via the dependence of coefficients $v_{3,5,8}$ on the confinement energy. This is only effective when the height is substantially affecting the confinement energy, i.e. for $H<R$. Although for small radius ${\boldsymbol \mu}_{\text{LC-EV}}$ can become larger than ${\boldsymbol \mu}_{\text{IC-BV}}$, the validity of the envelope function approximation starts to break down. This is particularly true for the envelope velocity related quantities, which depend on the gradient of the envelope wave functions.

Finally we point out that the spin moment is almost constant at one Bohr magneton, see Fig.~\ref{fig:j_mu_cylinder}(d). The reason is the same as was found for the spheres: although there is a sizeable intermixing of valence band states, the effect of different bands on the spin moment cancel to a large degree each other out. From Fig.~\ref{fig:j_mu_cylinder}(d) we observe that the radius dependence of the spin moment is different from the height dependence, i.e. the (small) corrections are different for a nanowire and quantum well. This shows once more that the radius and height have a different influence on the disks.

\subsection{Disks with soft boundaries} \label{sec:softdisks}

\begin{figure}[t]
\begin{center}
\includegraphics[width=\columnwidth]{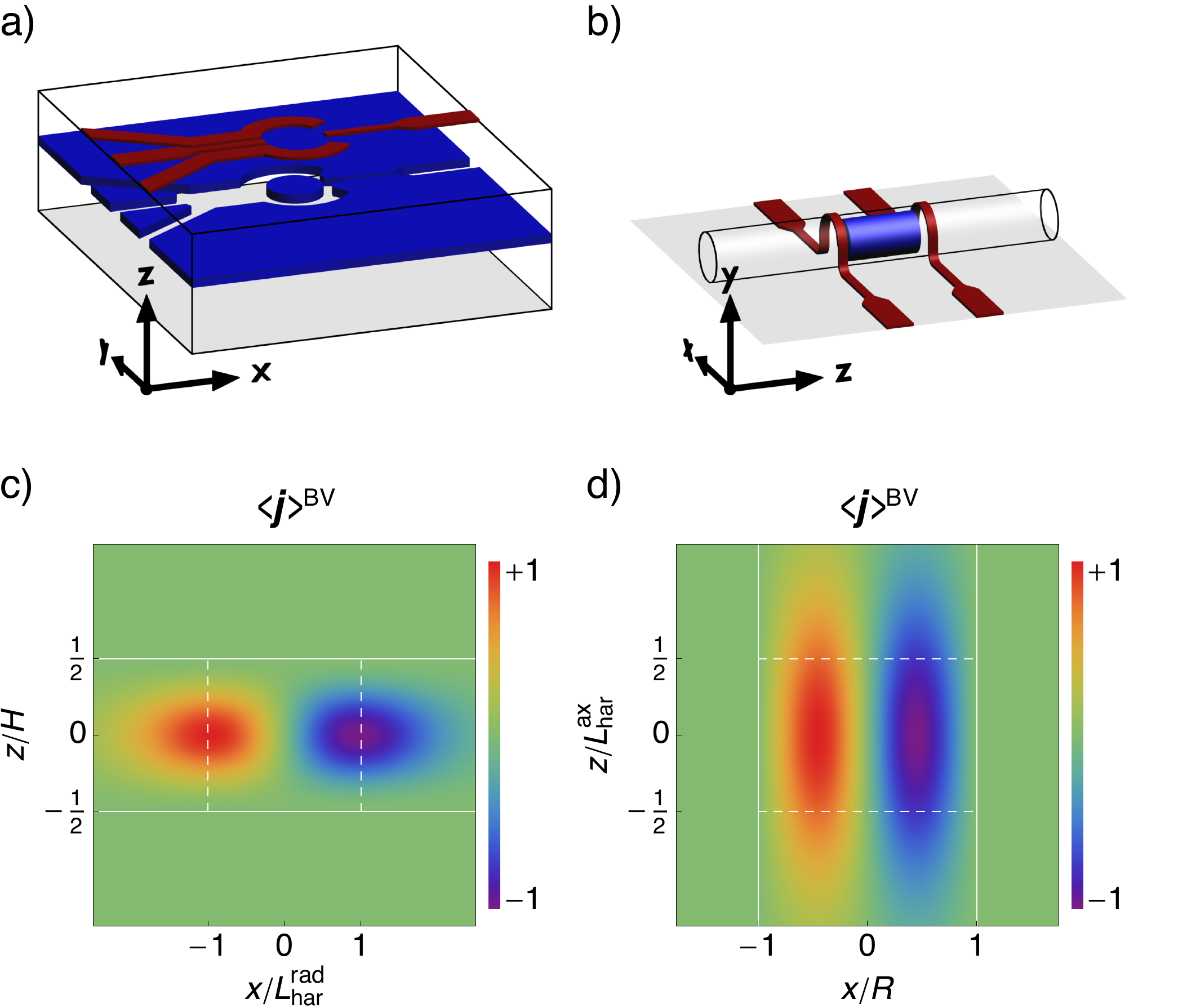}
\caption{(a) Gates (red) define by electrostatic means a quantum dot in a quantum well (blue). We take hard-wall confinement in the $z$-direction and a harmonic potential in the lateral direction. (b) Gates (red) define by electrostatic means a quantum dot in a nanowire (transparent). We take hard-wall confinement at the nanowire surface and a harmonic confinement potential in the axial direction. (c) The normalized magnitude of $\langle {\bf j} \rangle^{\text{BV}}$ in the ${\bf e}_y$-direction of an InAs quantum well with $H=10$~nm and $L_{\text{har}}^{\text{rad}}=10$~nm\cite{Hanson2007}. (d) The normalized magnitude of $\langle {\bf j} \rangle^{\text{BV}}$ in the ${\bf e}_y$-direction of an InAs nanowire with $R=40$~nm and $L_{\text{har}}^{\text{ax}}=10$~nm\cite{Csonka2008}. The continuous white lines indicate hard-wall boundaries, the dashed ones indicate the harmonic confinement length.}
\label{fig:j_harmonic}
\end{center}
\end{figure}

To show that the qualitative picture of the spin-orbit correlated currents and resulting moments does not depend on the choice of hard-wall boundaries, we will now investigate cylindrically symmetric nanostructures with soft boundaries. Such boundaries can arise when quantum dots are electrostatically defined using gates on quantum wells, see Fig.~\ref{fig:j_harmonic}(a). The confining potential of such a gate-defined quantum dot in a quantum well having height $H$ is given by:
\begin{eqnarray}
V(r,z) = \left\{
\begin{array}{ll}
\frac{1}{2}m_0 \omega^2 r^2 & |z| \leq \frac{1}{2}H \\
\infty & |z| > \frac{1}{2}H
\end{array}
\right.
\end{eqnarray}
where we take a hard-wall boundary in the $z$-direction and a harmonic potential in the lateral direction having an oscillator frequency $\omega$. Soft boundaries also arise in gate-defined quantum dot in a nanowires, see Fig.~\ref{fig:j_harmonic}(b), of which the confining potential can be described as:
\begin{eqnarray}
V(r,z) = \left\{
\begin{array}{ll}
\frac{1}{2}m_0 \omega^2 z^2 & r \leq R \\
\infty & r > R
\end{array}
\right.
\end{eqnarray}
where $R$ is the radius of the nanowire, and we take a hard-wall boundary at the nanowire surface and a harmonic potential in the axial direction.

Unfortunately the Schr{\"o}dinger equation cannot be solved analytically for either of these two confinement potentials when taking all eight bands into account. However, we can expand the electron ground state into free cylindrical waves $\Psi^{\text{free}}_{F_z,{\bf k}}({\bf r})$:
\begin{eqnarray}
\Psi({\bf r}) = \int_{V_{\bf k}} c({\bf k})\Psi^{\text{free}}_{+1/2,{\bf k}}({\bf r})~d^3k
\end{eqnarray}
where $c({\bf k})$ are expansion coefficients, and ${\bf k}=(k,k_z)$ is the wave vector of the free cylindrical wave. We have limited the expansion to $F_z=+\tfrac{1}{2}$ states, since we are only interested in the electron ground state of the quantum dots. A free cylindrical wave $\Psi^{\text{free}}_{F_z,{\bf k}}$ is straightforwardly described in terms of the product states $|F_z;J,J_z;k,k_z\rangle$ (see Sec.\ref{sec:cylindrical}):
\begin{eqnarray}
\Psi^{\text{free}}_{F_z,{\bf k}}({\bf r}) = \sum_{J,J_z} v_{J,J_z}({\bf k}) \langle {\bf r} |F_z;J,J_z;k,k_z\rangle
\end{eqnarray}
where coefficients $v_{J,J_z}({\bf k})$ follow from diagonalizing the Hamiltonian. These coefficients can one-to-one be identified with the (intermixing) coefficients $v_i$ of the disks.

\begin{figure}
\begin{center}
\includegraphics[width=\columnwidth]{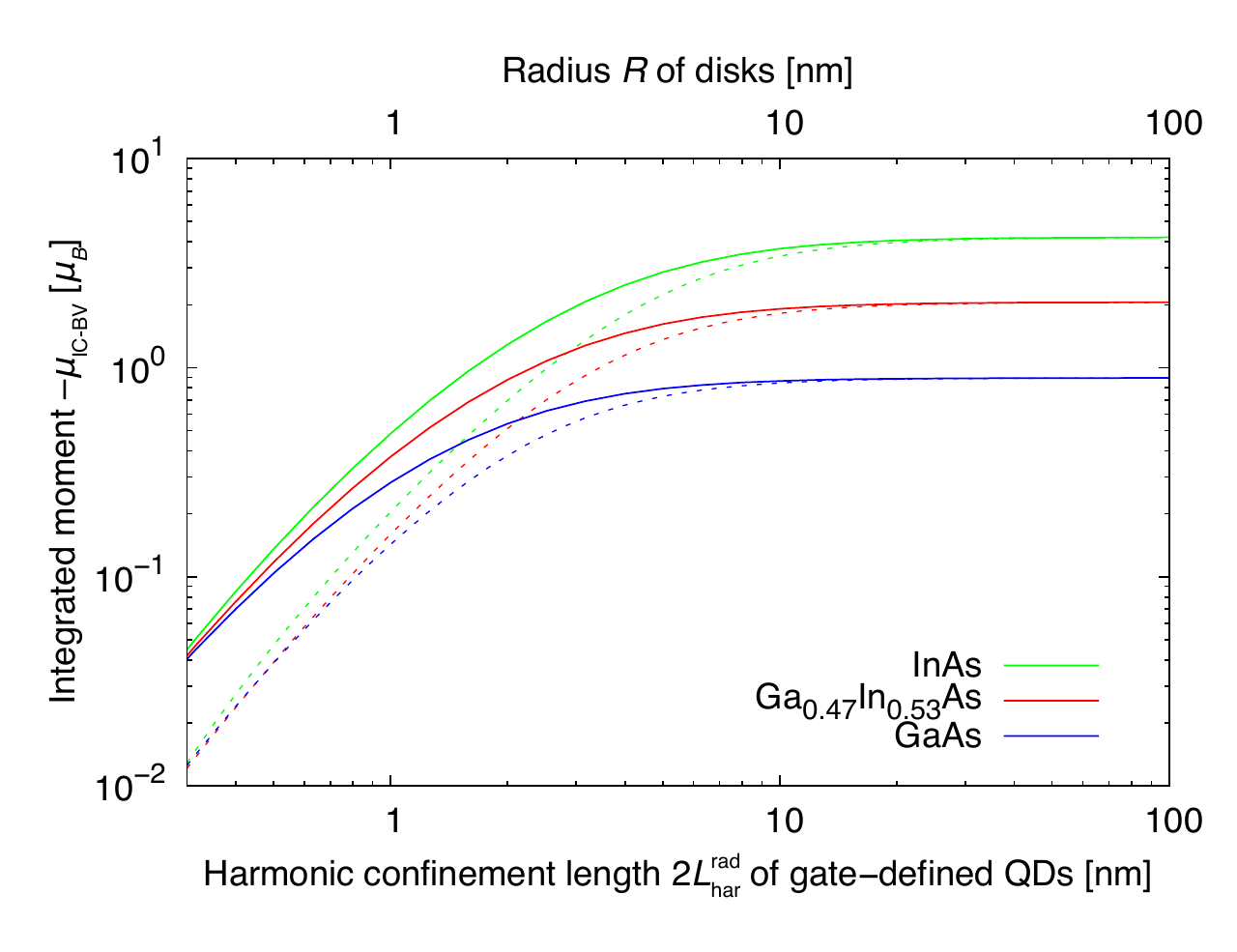}
\caption{The integrated orbital moment ${\boldsymbol \mu}_{\text{IC-BV}}$ as function of the harmonic confinement length $L_{\text{har}}^{\text{rad}}$ for gate-defined quantum dots in quantum wells (continuous lines) and as function of radius $R$ for disk (dotted lines), with $H=10$~nm.}
\label{fig:mu_harmonic_QW}
\end{center}
\end{figure}

Based on the analysis of the spheres and disks, we know that the dominant contribution to the electron ground state comes from the conduction band. We can therefore approximate the expansion coefficients using the envelope wave functions of a calculation involving only a single (conduction) band, $\Psi^{\text{single}}({\bf r})$:
\begin{eqnarray}
c({\bf k}) \approx \int_V \Psi^{\text{single}}({\bf r}) \Psi^{\text{free, single}}_{+1/2,{\bf k}}({\bf r})~d^3r
\end{eqnarray}
where the free cylindrical wave of a single (conduction) band is given by:
\begin{eqnarray}
\Psi^{\text{free, single}}_{+1/2,{\bf k}}({\bf r}) = \langle {\bf r} |F_z=+\tfrac{1}{2};J=\tfrac{1}{2},J_z=+\tfrac{1}{2};k,k_z\rangle\nonumber \\
\end{eqnarray}
We take for the envelope wave functions involving only the conduction band the solutions of Ref.~\onlinecite{Ikhdair2012}; for a quantum dot in a quantum well this is:
\begin{eqnarray}
\Psi^{\text{single}}_{\text{QW}}({\bf r}) = N e^{-\left(r/2L_{\text{har}}^{\text{rad}}\right)^2}\cos\left(\pi\frac{z}{H}\right)
\end{eqnarray}
where $L_{\text{har}}^{\text{rad}}=\sqrt{\hbar/2 m_0 \omega}$ is the harmonic confinement length in the lateral direction, and $N$ a normalization constant. Similarly,  for a quantum dot in a nanowire we have:
\begin{eqnarray}
\Psi^{\text{single}}_{\text{NW}}({\bf r}) = N e^{-\left(z/\sqrt{2}L_{\text{har}}^{\text{ax}}\right)^2}J_0\left(\rho_{0,1}\frac{r}{R}\right)
\end{eqnarray}
where $L_{\text{har}}^{\text{ax}}=\sqrt{\hbar/m_0 \omega}$ the harmonic confinement length in the axial direction. Using these single band envelope wave functions, we find an approximation for $\Psi({\bf r})$, of which the accuracy depends on the amount of intermixing of valence band states. Although it is possible to solve this problem analytically, for practical reasons we used only a limited number of free cylindrical waves in the expansion and calculated numerically $v_{J,J_z}({\bf k})$ for each wave. This numerical approximation converges when we use $\sim50-100$ free cylindrical waves. 

Now that the electron ground state is determined, we can use the techniques outlined in Sec.~\ref{sec:orbmu} to calculate the spin-orbit correlated currents. In Fig.~\ref{fig:j_harmonic}(c) we show $\langle {\bf j} \rangle^{\text{BV}}$ for a realistic gate-defined quantum dot in an InAs quantum well\cite{Hanson2007}. This current distribution is very similar to the one found for disks, see Fig.~\ref{fig:j_mu_cylinder}(a); the only difference is that the current is more smeared out in the lateral direction. In Fig.~\ref{fig:mu_harmonic_QW} we show the dependence of the integrated moment ${\boldsymbol \mu}_{\text{IC-BV}}$ on $L_{\text{har}}^{\text{rad}}$ for gated-defined quantum dots in quantum wells with $H=10$~nm. In the same graph, we show the radius dependence of disks having $H=10$~nm, so that we can directly observe the difference between hard-wall and soft boundaries. As expected, the boundaries have no influence in the limit of large $L_{\text{har}}^{\text{rad}}$ or $R$. When decreasing the size of the quantum dots, the quenching starts earlier for hard-wall boundaries than for the soft boundaries. This is easily understood by comparing the current distributions: the current is more smeared out in the lateral direction for the gate-defined quantum dots, meaning that they have a larger orbital moment for the same (effective) radius. At very small sizes, the rate of quenching is the same for hard-wall and soft boundaries. We therefore conclude that the net effect of the soft boundaries in the lateral direction is to merely change the onset of quenching, yet the underlying mechanisms remain the same.

\begin{figure}
\begin{center}
\includegraphics[width=\columnwidth]{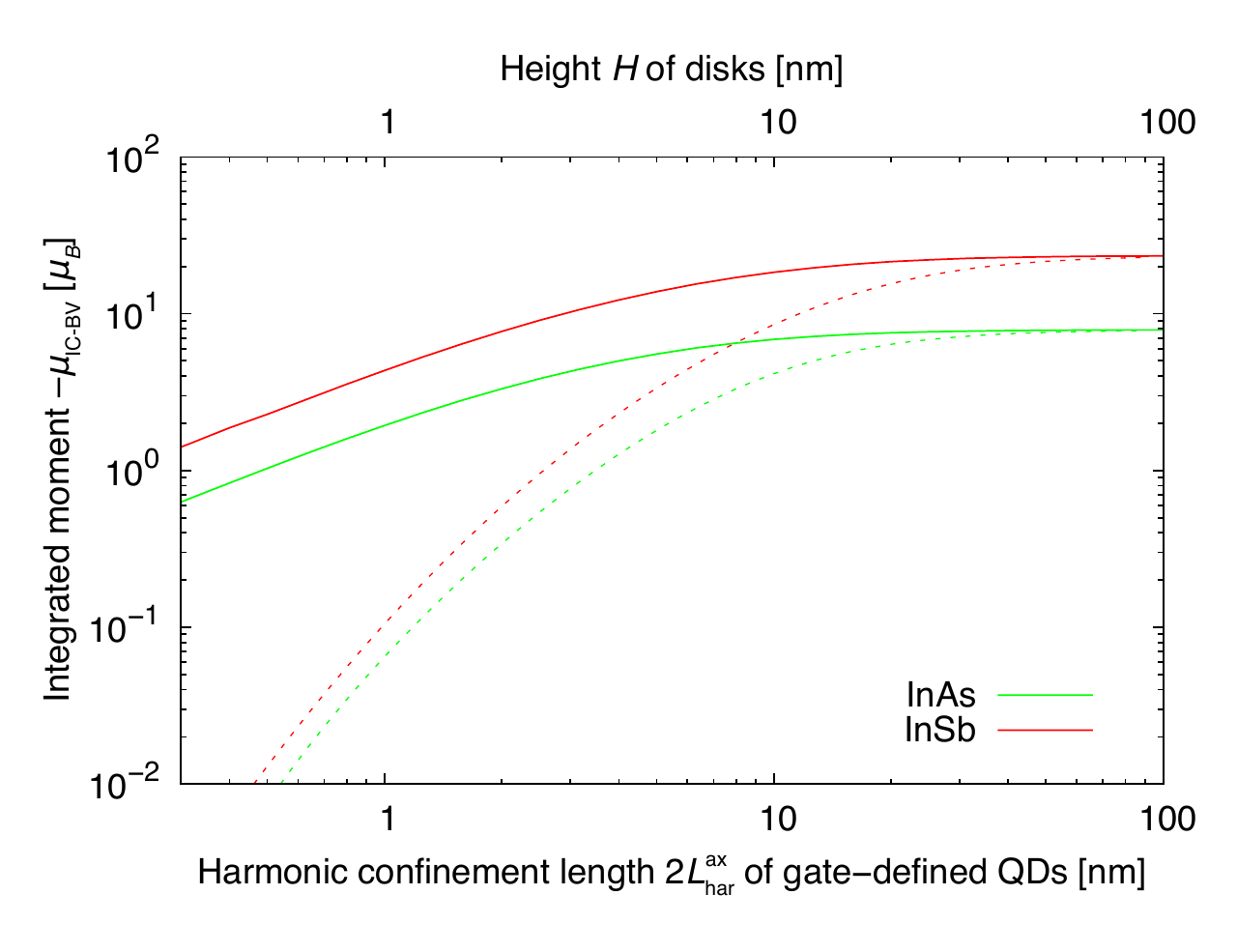}
\caption{The integrated orbital moment ${\boldsymbol \mu}_{\text{IC-BV}}$ as function of the harmonic confinement length $L_{\text{har}}^{\text{ax}}$ for gate-defined quantum dots in nanowires (continuous lines) and as function of height $H$ for disks (dotted lines), with $R=40$~nm\cite{Csonka2008,Nilsson2009}.}
\label{fig:mu_harmonic_NW}
\end{center}
\end{figure}

The current distribution for a gate-defined quantum dot in an InAs nanowire\cite{Csonka2008} is shown in Fig.~\ref{fig:j_harmonic}(d). In this case the current is smeared out in the axial direction, when comparing it to the disks, see Fig.~\ref{fig:j_mu_cylinder}(a). The dependence of the orbital moment ${\boldsymbol \mu}_{\text{IC-BV}}$ on $L_{\text{har}}^{\text{ax}}$ is plotted in Fig.~\ref{fig:mu_harmonic_NW}, along with the height dependence of the corresponding disks. Although the orbital moment of hard-wall and soft boundaries is again the same for large quantum dots, the rate of quenching at small sizes is different: for the disks the rate is proportional to $H^4$, while for the gate-defined quantum dots the rate is proportional to $L_{\text{har}}^{\text{ax}}$. We do not understand this difference, since the current distributions are qualitatively the same. It could result from the approximation scheme we have used to retrieve the electron ground state. To investigate such unwanted effects, a direct numerical calculation of the electron ground state would be needed.

\subsection{Rings} \label{sec:rings}

\begin{figure}
\begin{center}
\includegraphics[width=\columnwidth]{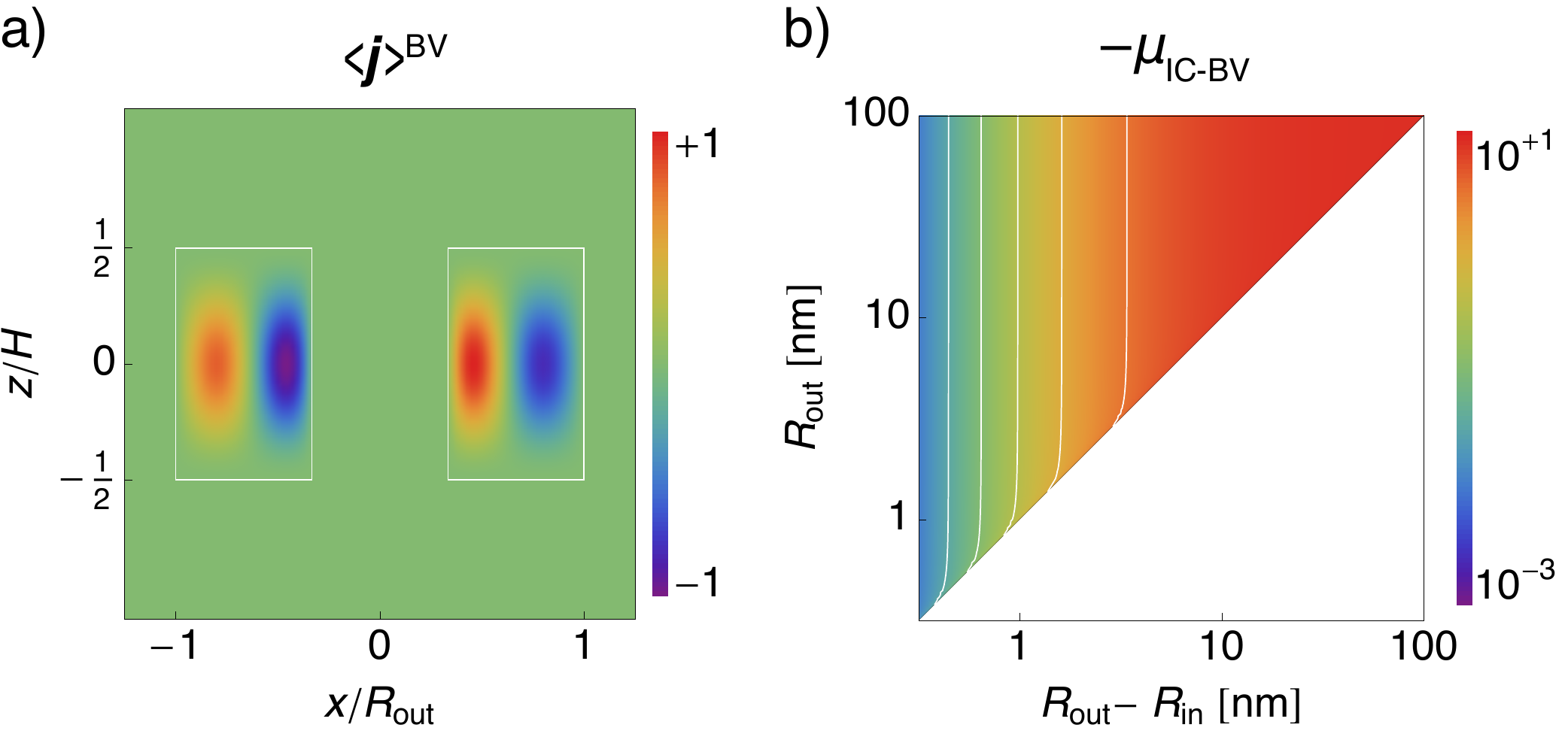}
\caption{(a)~The $xz$-cross-section of the spatial distribution of the normalized magnitude of the ${\bf e}_{y}$-component of $\langle {\bf j}\rangle^{\text{BV}}$ of a ring. Similar to the spherical shells, there are two oppositely circulating current loops. For the plot we choose $R_{\text{in}}/R_{\text{out}}=\tfrac{1}{3}$ and set $v_{6}=0$. (b)~The dependence of ${\boldsymbol \mu}_{\text{IC-BV}}$ (in $\mu_B$) on the ring thickness $R_{\text{out}}-R_{\text{in}}$ and outer radius $R_{\text{out}}$, for an InAs ring with $H=100$~nm. Similar to the spherical shells, the orbital moment depends only on the ring thickness, as can be seen from the white contour lines. We have set $v_6=0$ to avoid numerical artifacts in the calculation.}
\label{fig:j_mu_ring}
\end{center}
\end{figure}

In Sec.~\ref{sec:harddisks} we found that lowering the symmetry from spherical to cylindrical, one more independent handle on the magnetic moment is introduced. Analogous to the analysis of spherical shells, it proves interesting to see what effect the topology has on cylindrically symmetric nanostructures. Moreover, the removal of material from the center of the spheres lead to new currents, and such effects might now be expected for cylindrical nanostructures too. We will therefore analyze a ring, with inner radius $R_{\text{in}}$, outer radius $R_{\text{out}}$, and height $H$, of which the confining potential is given by:
\begin{eqnarray}
V(r,z) = \left\{
\begin{array}{lc}
0 & R_{\text{in}} \leq r \leq R_{\text{out}} \quad \text{and} \quad |z|\leq H/2 \\
\infty & \text{elsewhere}
\end{array}
\right.
\end{eqnarray}
The Neumann functions $N_{L_{\text{E},z}}(r)$ do play a role now, since the origin is not involved in the wave function (see Sec.~\ref{sec:cylindrical}). The parameter $\xi$ is therefore non-zero and should follow from the boundary conditions. Since the electron ground state predominantly originates from conduction band states, we choose the approximate boundary condition:
\begin{eqnarray}
\langle r,\theta,z |k,k_z,0\rangle_{r=R_{\text{in}},r=R_{\text{out}},z=\pm\tfrac{H}{2}}=0
\end{eqnarray}
This condition leads to the system of equations:
\begin{eqnarray}
\left\{
\begin{array}{r}
J_0(k R_{\text{in}}) + \xi N_0(k R_{\text{in}}) = 0 \\
J_0(k R_{\text{out}}) + \xi N_0(k R_{\text{out}}) = 0
\end{array}
\right.
\end{eqnarray}
which determine $(\xi,k)$ for a given $(R_{\text{in}},R_{\text{out}})$. Although this system of equations is not generally analytically solvable, it can be inferred that both $\xi$ and $kR_{\text{out}}$ depend only on the ratio $R_{\text{in}}/R_{\text{out}}$. This can also be seen when analyzing the asymptotic limit of the equations, which results in approximate solutions:
\begin{eqnarray}
k &\approx& \frac{\pi}{R_{\text{out}} - R_{\text{in}}} \\
\xi &\approx& \tan\left(k R_{\text{out}} + \tfrac{\pi}{4}\right)
\end{eqnarray} 
These approximate relations resemble the ones found for the spherical shells. In Fig.~\ref{fig:j_mu_ring}(a) we plot the current distribution of a ring with $R_{\text{in}}/R_{\text{out}}=\tfrac{1}{3}$. Analogous to the spherical shells, the existence of the inner surface leads to an additional oppositely circulating current. This shows once more that the topology  of the nanostructure has a profound influence on the orbital current distribution. Contrary to the spherical shells, we find that these two current loops carry an equal amount of current, so that, irrespective of the size, the integrated current is zero. The orbital moments generated by each of the currents will partially cancel, the degree of cancellation depending on the ring thickness. This result was to be expected, since the radial wave number is determined by the ring thickness $R_{\text{out}}-R_{\text{in}}$, and the orbital moment of a disk depends on $R/\rho_{0,1}=1/k$. The orbital moment can therefore only be tuned either via the thickness or the height of the ring. It seems, therefore, that changing the topology of the nanostructure does not generate additional handles on the orbital moment, while changing the spatial symmetry does have this effect.

To be complete, the above reasoning only holds as long as the approximate solution is valid: in general, $k$ might not depend only on the ring thickness. We have therefore computed numerically the solution of the boundary conditions, and used them to numerically calculate the radius dependence of the most important integrated orbital moment ${\boldsymbol \mu}_{\text{IC-BV}}$, see Fig.~\ref{fig:j_mu_ring}(b). It can readily be seen that this orbital moment depends only on the ring thickness. Only when $R_{\text{in}}$ approaches zero, $R_{\text{out}}$ starts to have an influence too. These small values of $R_{\text{in}}$ correspond to an inner region comparable to the unit cell of the crystal, and the validity of the envelope function approximation is questionable. Finally we note that having a finite barrier will also lead to substantial changes at small $R_{\text{in}}$: tunneling through the inner region will reduces the strength of the inner current loop and decreases the radius of the outer current loop, both leading to a reduction of the degree of cancellation of the orbital moments.

\section{Conclusions} \label{sec:con}

We have found that the origin of spin-correlated currents of different nanostructures is related to the intermixing of valence band states into the electron ground state. Irrespective of the geometrical symmetry (spherical vs. cylindrical), type of boundaries (hard-wall vs. soft), and material, we have found that the dominant current circulates within the nanostructure, peaking roughly halfway between the center and edge of the nanostructure. This distribution can be regarded as a simple current loop, which generates the orbital moment. By changing the size of the nanostructure, both the amount of current (intermixing of valence states) and the lever arm are changed, leading to quenching of the orbital moment for small sizes. For spherically symmetric nanostructures we have found that the orbital moment and confinement energy are parameterized by a single geometrical parameter. By lowering the symmetry, such as for cylindrically symmetric nanostructures, we have found that these two quantities can be independently tuned: the radius and height have different influences on disks. Although changing the topology of nanostructures can introduce an additional geometrical handle on the orbital moment, we have observed that the orbital moment and confinement energy are then parameterized by a combination of geometrical parameters. Such handles can be interesting in relation to tuning the orbital moment, i.e. manipulating the $g$ tensor, for active manipulation of the electron spin.

\bibliography{central-bibliography}

\begin{sidewaystable}
\begin{center}

\vspace{60mm}
\begin{footnotesize}
\caption{The cylindrically symmetrical Hamiltonian ${\cal H}_{\text{cyl}}$ represented in a cylindrical envelope basis~\citep{Trebin1979,Sercel1990}. We have used the abbreviations: {\footnotesize $E_{\text{CB}}= \frac{\hbar ^2 \left(k^2+k_z^2\right)}{2 m_0}$, $E_{\text{HH}}= E_g + \frac{\hbar^2 k^2}{2m_0}\left(\gamma_1 + \gamma_2\right) + \frac{\hbar^2 k_z^2}{2m_0}\left(\gamma_1 - 2\gamma_2\right)$, $E_{\text{LH}}= E_g + \frac{\hbar^2 k^2}{2m_0}\left(\gamma_1 - \gamma_2\right) + \frac{\hbar^2 k_z^2}{2m_0}\left(\gamma_1 + 2\gamma_2\right)$, $E_{\text{SO}}= E_g + \Delta + \frac{\hbar^2\left(k^2 + k_z^2\right)}{2m_0} \gamma_1$}} \label{table:Hcyl}
\begin{tabular}{>{$}c<{$}>{$}c<{$}>{$}c<{$}>{$}c<{$}>{$}c<{$}>{$}c<{$}>{$}c<{$}>{$}c<{$}}
\hline \hline
  \text{CB}\uparrow & \text{CB}\downarrow & \text{HH}\uparrow & \text{LH}\uparrow & \text{LH}\downarrow & \text{HH}\downarrow & \text{SO}\uparrow & \text{SO}\downarrow \\
\hline
  \\
  E_{\text{CB}}               &  0                            & -i\sqrt{\frac{1}{2}} k P_0                                         &  i\sqrt{\frac{2}{3}} k_z P_0                                               &  i\sqrt{\frac{1}{6}} k   P_0                                               & 0                                                                          & -i\sqrt{\frac{1}{3}} k_z P_0                                               & -i\sqrt{\frac{1}{3}} k   P_0 \\
  0                           &  E_{\text{CB}}                & 0                                                                  & -i\sqrt{\frac{1}{6}} k   P_0                                               &  i\sqrt{\frac{2}{3}} k_z P_0                                               & i\sqrt{\frac{1}{2}} k P_0                                                  & -i\sqrt{\frac{1}{3}} k   P_0                                               &  i\sqrt{\frac{1}{3}} k_z P_0 \\
  i\sqrt{\frac{1}{2}} k   P_0 &  0                            & - E_{\text{HH}}                                                    &  \sqrt{3} \frac{\hbar^2 k k_z}{m_0}\gamma_3                                &  \sqrt{3} \frac{\hbar^2 k^2}{2 m_0} \tfrac{1}{2}(\gamma_2+\gamma_3)        & 0                                                                          & -\sqrt{\frac{3}{2}} \frac{\hbar^2 k k_z}{m_0} \gamma_3                     & -\sqrt{\frac{3}{2}} \frac{\hbar^2 k^2}{2 m_0}(\gamma_2+\gamma_3) \\
 -i\sqrt{\frac{2}{3}} k_z P_0 &  i \sqrt{\frac{1}{6}} k   P_0 & \sqrt{3} \frac{\hbar^2 k k_z}{m_0} \gamma_3                        & - E_{\text{LH}}                                                            &   0                                                                        &  \sqrt{3} \frac{\hbar^2 k^2}{2 m_0} \tfrac{1}{2}(\gamma_2+\gamma_3)        & -\sqrt{\frac{1}{2}} \frac{\hbar^2 \left(k^2 - 2k_z^2\right)}{m_0} \gamma_2 &  \sqrt{\frac{9}{2}} \frac{\hbar^2 k k_z}{m_0} \gamma_3 \\
 -i\sqrt{\frac{1}{6}} k   P_0 & -i \sqrt{\frac{2}{3}} k_z P_0 & \sqrt{3} \frac{\hbar^2 k^2}{2 m_0} \tfrac{1}{2}(\gamma_2+\gamma_3) & 0                                                                          & - E_{\text{LH}}                                                            & -\sqrt{3} \frac{\hbar^2 k k_z}{m_0}\gamma_3                                &  \sqrt{\frac{9}{2}} \frac{\hbar^2 k k_z}{m_0} \gamma_3                     &  \sqrt{\frac{1}{2}} \frac{\hbar^2 \left(k^2 - 2k_z^2\right)}{m_0} \gamma_2 \\
  0                           & -i \sqrt{\frac{1}{2}} k   P_0 & 0                                                                  &  \sqrt{3} \frac{\hbar^2 k^2}{2 m_0} \tfrac{1}{2}(\gamma_2+\gamma_3)        & -\sqrt{3} \frac{\hbar^2 k k_z}{m_0} \gamma_3                               & - E_{\text{HH}}                                                            &  \sqrt{\frac{3}{2}} \frac{\hbar^2 k^2}{2 m_0} (\gamma_2+\gamma_3)          & -\sqrt{\frac{3}{2}} \frac{\hbar^2 k k_z}{m_0} \gamma_3 \\
  i\sqrt{\frac{1}{3}} k_z P_0 &  i \sqrt{\frac{1}{3}} k   P_0 & -\sqrt{\frac{3}{2}} \frac{\hbar^2 k k_z}{m_0} \gamma_3             & -\sqrt{\frac{1}{2}} \frac{\hbar^2 \left(k^2 - 2k_z^2\right)}{m_0} \gamma_2 &  \sqrt{\frac{9}{2}} \frac{\hbar^2 k k_z}{m_0} \gamma_3                     &  \sqrt{\frac{3}{2}} \frac{\hbar^2 k^2}{2 m_0} (\gamma_2+\gamma_3)          & - E_{\text{SO}}                                                            & 0 \\
  i\sqrt{\frac{1}{3}} k   P_0 & -i \sqrt{\frac{1}{3}} k_z P_0 & -\sqrt{\frac{3}{2}} \frac{\hbar^2 k^2}{2 m_0}(\gamma_2+\gamma_3)   &  \sqrt{\frac{9}{2}} \frac{\hbar^2 k k_z}{m_0} \gamma_3                     &  \sqrt{\frac{1}{2}} \frac{\hbar^2 \left(k^2 - 2k_z^2\right)}{m_0} \gamma_2 & -\sqrt{\frac{3}{2}} \frac{\hbar^2 k k_z}{m_0} \gamma_3                     & 0                                                                          & - E_{\text{SO}} \\
\hline \hline
\end{tabular}
\end{footnotesize}

\vspace{15mm}
\caption{The definition of the Bloch states (similar to Refs.~\onlinecite{Sercel1990}~or~\onlinecite{Trebin1979}).} \label{table:Bloch}
\begin{tabular}{r>{$}c<{$}>{$}l<{$}r>{$}c<{$}>{$}l<{$}}
\hline \hline 
CB$\uparrow$   &=& |s\rangle\uparrow   & LH$\uparrow$   & = & -\sqrt{\frac{1}{6}} \left[|x\rangle + i |y\rangle\right]\downarrow + \sqrt{\frac{2}{3}} |z\rangle \uparrow \\
CB$\downarrow$ &=& |s\rangle\downarrow & LH$\downarrow$ & = & +\sqrt{\frac{1}{6}} \left[|x\rangle - i |y\rangle\right]\uparrow   + \sqrt{\frac{2}{3}} |z\rangle \downarrow  \\
HH$\uparrow$   & = & -\sqrt{\frac{1}{2}} \left[|x\rangle + i |y\rangle\right]\uparrow   & SO$\uparrow$   &=& -\sqrt{\frac{1}{3}} \left[|x\rangle + i |y\rangle\right]\downarrow - \sqrt{\frac{1}{3}} |z\rangle \uparrow   \\
HH$\downarrow$ & = & +\sqrt{\frac{1}{2}} \left[|x\rangle - i |y\rangle\right]\downarrow & SO$\downarrow$ &=& -\sqrt{\frac{1}{3}} \left[|x\rangle - i |y\rangle\right]\uparrow   + \sqrt{\frac{1}{3}} |z\rangle \downarrow \\
\hline \hline
\end{tabular}

\end{center}
\end{sidewaystable}

\end{document}